\documentclass[sigconf, acmsmall, nonacm]{acmart}
\usepackage{xspace}
\usepackage{multirow} 
\usepackage{cleveref}
\usepackage{subcaption}
\newfloat{floatquote}{tbp}{loq}
\floatname{floatquote}{Supplementary Text}
\crefname{floatquote}{quote}{quotes}
\Crefname{floatquote}{Supplementary Text}{Quotes}
\newcommand{\proto}{\textit{DesignMinds}\xspace}
\AtBeginDocument{%
  }

\setcopyright{acmlicensed}
\copyrightyear{2018}
\acmYear{2018}
\acmDOI{XXXXXXX.XXXXXXX}

\acmConference[Conference acronym 'XX]{Make sure to enter the correct conference title from your rights confirmation email}{June 03--05, 2018}{Woodstock, NY}
\acmISBN{978-1-4503-XXXX-X/18/06}

\begin{document}

\title{DesignMinds: Enhancing Video-Based Design Ideation with Vision-Language Model and Context-Injected Large Language Model}

\author{Tianhao He}
\affiliation{%
  \institution{Delft University of Technology}
  \streetaddress{Landbergstraat 15}
  \city{Delft}
  \postcode{2628 CE}
  \country{The Netherlands}}
\email{t.he-1@tudelft.nl}

\author{Andrija Stanković}
\affiliation{%
  \institution{Delft University of Technology}
  \streetaddress{Landbergstraat 15}
  \city{Delft}
  \country{Netherlands}}
\email{andrija.stkvc@gmail.com}

\author{Evangelos Niforatos}
\affiliation{%
  \institution{Delft University of Technology}
  \streetaddress{Landbergstraat 15}
  \city{Delft}
  \country{Netherlands}}
\email{e.niforatos@tudelft.nl}

\author{Gerd Kortuem}
\affiliation{%
  \institution{Delft University of Technology}
  \streetaddress{Landbergstraat 15}
  \city{Delft}
  \postcode{2628 CE}
  \country{The Netherlands}}
\email{g.w.kortuem@tudelft.nl}
\renewcommand{\shortauthors}{He et al.}

\begin{abstract}
Ideation is a critical component of video-based design (VBD), where videos serve as the primary medium for design exploration and inspiration. The emergence of generative AI offers considerable potential to enhance this process by streamlining video analysis and facilitating idea generation. In this paper, we present \proto, a prototype that integrates a state-of-the-art Vision-Language Model (VLM) with a context-enhanced Large Language Model (LLM) to support ideation in VBD. To evaluate \proto, we conducted a between-subject study with 35 design practitioners, comparing its performance to a baseline condition. Our results demonstrate that \proto significantly enhances the flexibility and originality of ideation, while also increasing task engagement. Importantly, the introduction of this technology did not negatively impact user experience, technology acceptance, or usability.

\end{abstract}



\begin{teaserfigure}
\centering
  \includegraphics[width=.5\textwidth]{ 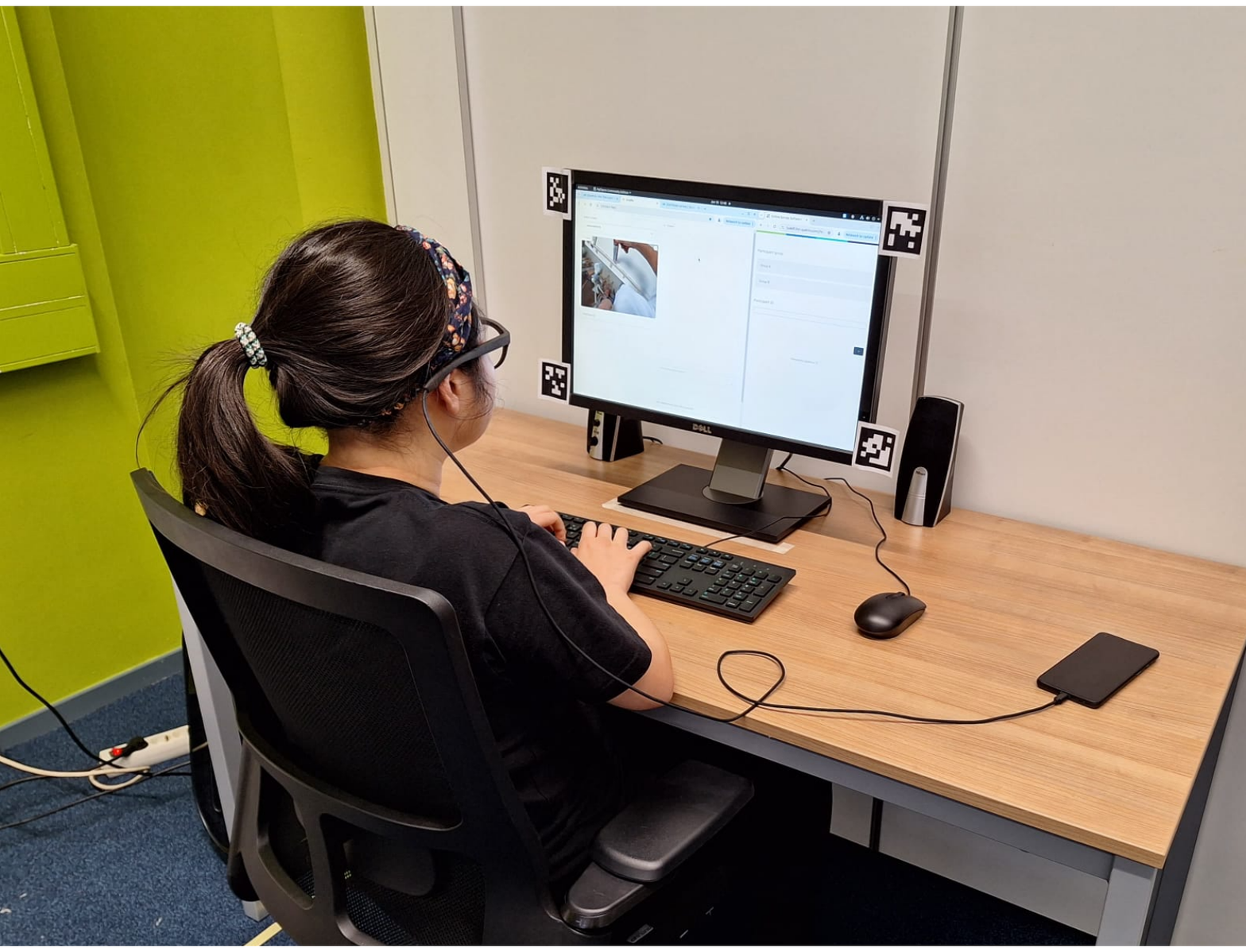}
  \caption{
A designer in the experimental group is interacting with \proto.}
  \label{fig:studyScence}
  \Description{A designer in the experimental group is interacting with DesignMinds. The image shows a person sitting in front of a computer at a desk. The designer is wearing glasses and has their hair tied back. They are focused on a desktop monitor, interacting with software. The person is using a mouse and keyboard, and the computer screen displays a digital interface. The screen features a window with images and text, and there are several QR code markers attached to the monitor, used for calibration and tracking purposes. A phone is placed on the desk next to the keyboard. The environment appears to be a typical research lab setup.}
\end{teaserfigure}

\maketitle

\section{Introduction}

Idea generation is the cornerstone of innovation and serves as the foundation for new designs \cite{lopez-mesa_effects_2011, chulvi_comparison_2012}. Video-Based Design (VBD) enables designers to utilize video content as a key tool for generating knowledge, inspiring new ideas, and identifying potential challenges \cite{jacob_design_2007, tatar_1989, vertelney_1989, Fucs_2020}. The ideation of VBD plays a crucial role in brainstorming to produce a wide range of ideas, which are then filtered and refined to develop optimal solutions \cite{mackay_video_1999, mackay_video_2000, dam2017ideation}. However, generating novel design ideas from videos is challenging for a large group of practitioners. It requires not only a significant investment of time and effort but also extensive design experience to generate a substantial number of related ideas for practice \cite{dam2017ideation}. Consolidating design problems and generating feasible solutions from videos using traditional VBD methods typically requires extensive video review and the application of professional divergent thinking \cite{jacob_design_2007}. This process is often labor-intensive and heavily dependent on the practitioner's design experience and knowledge, which can be particularly challenging for novice designers with limited expertise and resources \cite{1541974}.  Additionally, previous research indicated that advanced video tools can potentially enhance the design work with videos to improve the quality outcomes, and to facilitate interactions \cite{zahn_comparing_2010}. 

With the recent surge in Generative AI (GenAI), technologies such as the Large Language Model (LLM) GPT-4 \cite{openai2023gpt4} demonstrate significant potential to enhance creative tasks across various design domains. A base LLM model can generate ideas across diverse scopes. Its capabilities can be further refined by incorporating contextual material through a process known as Retrieval-Augmented Generation (RAG) to make it adaptable in current circumstances \cite{lewis2020retr}. Additionally, Vision-Language Models (VLMs) possess the ability to interpret videos with high detail, reducing the need for extensive human effort \cite{bordes_introduction_2024}. These advancements have the potential to assist designers in overcoming challenges associated with generating efficient and effective ideas, particularly when faced with prolonged video viewing and limited design experience \cite{rezwana_designing_2023, kim_tool_2021}. As such, this paper explores an approach that combines a customized VLM and LLM (\proto) to enhance the "watch-summarize-ideate" process in VBD tasks through designer-AI co-ideation. We then present our benchmarks and evaluate the quality of the generated ideas, cognitive processes, user experience (UX) and technology acceptance and use from VBD ideation. Our work makes the following contributions:
\begin{itemize}
\item We introduce a novel GenAI-powered chatbot that features video understanding and design-context-based idea recommendations to enhance the ideation capabilities of new VBD practitioners.
\item We investigate the impact of our prototype in terms of ideation quality, cognitive processing during ideation, and subsequent UX and technology acceptance.
\item Ultimately, we propose a potential tool (\proto) involving the use of a customized VLM and LLM to scale up the VBD ideation process for new designers.
\end{itemize}
Finally, our findings indicate that \proto improves the flexibility and originality of design ideas and boost design task engagement. The adoption of this technology also did not adversely affect the established patterns of UX, technology acceptance and usability.

\section{Background}
\subsection{Ideation in Design}
In the design process, ideation is a key aspect of experience that influences both the initiation and progression in the early stage of creative activities. Eckert and Stacey articulated that ideation is not merely a catalyst for creativity but also a critical component in developing design ideas \cite{eckert_sources_2000}. They claimed that ideation in design provides a contextual framework that enables designers to effectively communicate and position their work. It sparks design creativity, offering new perspectives and triggering the generation of original ideas  \cite{eckert_sources_2000}. Similarly, Setchi and Bouchard define ideation as a multifaceted phenomenon where designers absorb and reinterpret existing ideas, forms, and concepts \cite{setchi_search_2010}. This process is influenced by designers' individual experiences, cultural backgrounds, and personal interests and serves as a guiding principle for creativity. The subjectivity of ideation accelerates designers to explore a broader array of possibilities. Gonçalves et al. extended the understanding of ideation into later stages, asserting that designers maintain a limited range of external stimuli preferences. Both design students and professionals often favor visual stimuli such as images, objects, and video sources to encourage creativity \cite{goncalves_inspiration_2016}. 

However, relying on specific stimuli and designers' own knowledge may cause the risk of design fixation \cite{jansson_design_1991}. This phenomenon occurs when designers over-rely on specific knowledge directly associated with a problem or themselves during ideation, eventually inhibiting the design outcome \cite{youmans_design_2014, linsey_study_2010}. Viswanathan and Linsey claimed that the problem of fixation is pervasive and varies inversely with the level of design expertise. They suggested that it is especially prominent among novice designers, who tend to rely heavily on their predominant knowledge during ideation \cite{viswanathan_design_2013}. In addition, novice designers often struggle to analyze problems comprehensively and have difficulty seeking helpful information during ideation \cite{eastman_new_2001, cross_expertise_2004}. This phenomenon often leads to failures in framing problems and directing the search for solutions, ultimately diminishing the design outcome. Gonçalves pointed out that the lack of reflection in ideation could be addressed by developing computational tools to help designers efficiently find relevant stimuli. Such tools could assist inexperienced designers in exploring ideas that are semantically distant from the problem domain and expand space for ideation \cite{goncalves_inspiration_2016}. Similarly, the study by Dazkir et al. showed that while self-selected contexts in designers led to greater interest in the topic, they often failed to develop effective design solutions. This indicates that, although some autonomy is beneficial for developing design ideas, many inexperienced designers still need external intervention in the early stages to aid in ideation \cite{dazkir_exploration_2013}. As such, designers, especially those with limited experience, often need additional help and guidance from outside sources to enhance ideation.

\subsection{Videos for Design Ideation}
The use of video as a central tool for ideation, known as VBD, involves capturing information and analyzing solutions in design process. This technique is particularly prevalent in fields like user experience UX design, interaction design, and ethnographic research \cite{jacob_design_2007}. By recording user interactions with products or environments, videos provide a dynamic and context-rich data source for designers. Design videotapes are informative for practitioners to deepen context understandings and generate follow-up interventions \cite{jacob_design_2007, jacob_edit_2007}. Designer from the Apple Inc. utilized videos to envision new user interfaces (UIs) for their future computers \cite{vertelney_1989}. They utilized videos to benchmark new UIs and study user behavioral reactions through videotapes. In the same year, Tatar from PARC explored learning from repeated video observations of user behavior through stationary camera recordings and aimed to minimize erroneous assumptions in software development \cite{tatar_1989}. Tatar also emphasized the important role of using videos for ideation to pinpoint design solutions. Similarly, Ylirisku and Buur conceptualized the practice and highlighted that using videos for design ideation is instrumental for practitioners. Videos are an effective tool for learning from target users’ daily experiences and augment designers generate an abundance of ideas for design artifacts \cite{jacob_edit_2007}. Moreover, designers can ideate from the "thick descriptions" that videos capture about users' movements, interactions, and emotional transitions, which help in constructing design narratives and encapsulating individual thoughts. While video-based design idea generation presents significant opportunities, videos often contain complex content and frequent events \cite{jacob_design_2007, jacob_edit_2007}. The process of watching these videos can be labor-intensive and time-consuming. Videos with rich details and rapid sequences require viewers substantial information processing effort to analyze perceived information. As a result, designer may suffers risks of diminishing decision-making capability and result in a decline in ideation effectiveness \cite{campbell_task_1988, paquette_effect_1988}. Therefore, it is essential to develop strategies to mitigate fatigue and reduce the information processing effort for designers who use videos for inspiration, while ensuring that they retain the valuable information presented in videos.

\subsection{GenAI for Design Ideation}

Recent advancements in GenAI are driving significant changes across multiple disciplines. Large Language Models (LLMs), such as GPT-4 \cite{openai2023gpt4}, have shown remarkable capabilities in assisting creative tasks for design purposes \cite{zhou_how_nodate}. Xu et al. proposed an LLM-augmented framework that uses LLM prompts to generate unified cognition for practitioners and optimize the creative design process in a professional product design \cite{xu_llm_2024}. Another group of researchers proposed Jamplate, a protocol that leverages formatted prompts in LLMs to guide novice designers in real-time. This approach enhances their critical thinking and improves idea generation more effectively \cite{xu_jamplate_2024}. Makatura et al. explored the use of GPT-4 to generate textual design language and spatial coordinates for product design and adaptation in industry \cite{makatura_how_2023}. They highlighted that GPT-4’s reasoning capabilities offer significant value in novel design domains. When designers are inexperienced with a particular domain or working on a novel problem, GPT-4 can synthesize information from related areas to provide suitable advice. In addition, by extending LLMs with visual understanding capabilities, VLMs demonstrates a promising advancement for completing open-ended visual tasks using information extracted from videos \cite{pmlr-v202-li23q}. Moreover, many researchers recently made attempts to evolve VLMs into more complex and context-aware systems. For example, Zhou et al. introduced NavGPT, an LLM-based navigation agent that uses visual cues detected by a VLM to provide indoor navigation suggestions \cite{zhou_navgpt_2023}. They demonstrated that their system can generate high-level navigational suggestions from automatic observations and moving histories. Moreover, Picard et al. explored the use of GPT-4V(ison) \cite{openai2023gpt4}, a version of GPT-4 with vision-language capabilities, in product design. They investigated its application in design tasks, such as analyzing handwritten sketches and providing follow-up suggestions for material selection, drawing analysis, and spatial optimization. Their findings demonstrated that this LVM model can handle complex design idea generation with proficiency \cite{picard_concept_2023}. 

The merging of LLMs and VLMs presents an opportunity to enhance design, prompting us to explore this integration in the refined field of VBD. We are curious on if combining LLMs and VLMs can benefit VBD practitioners in their idea generations. To investigate, \textbf{we prototyped \proto that integrates a state-of-the-art (SOTA) VLM and LLM model with a context-injection technique. We conducted a study involving two video-based design tasks to assess the impact on design ideation, focusing on ideation quality, cognitive processes, user experience, and technology acceptance}.


\section{Our \proto Prototype}

\begin{figure}
  \centering
    \includegraphics[width=0.85\linewidth]{ 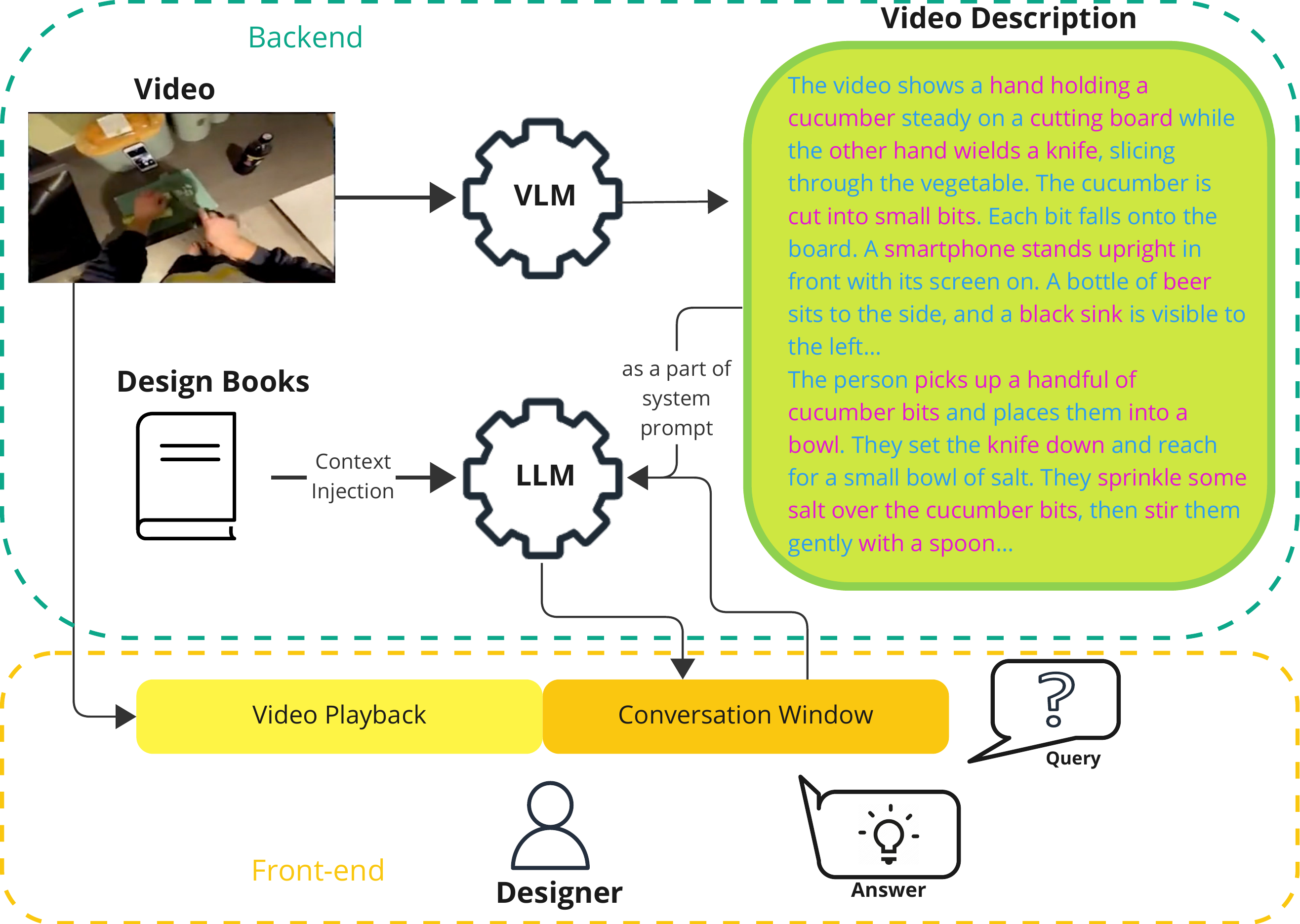}
    \caption{\proto consists of two primary components: the backend and the front-end. The backend includes a VLM and a LLM integrated with a design knowledge repository. The front-end features a video playback region alongside a conversational window. The videos are first processed to extract key terms (highlighted in pink in video description) and are then connected into a comprehensive description (blue in video description) using in-built language linking functions. These complete descriptions are then passed to the LLM, along with a knowledge repository enriched by selected design books from a committee vote. Designers can then use the features in front-end to watch the video playback to enhance trust and grounding for the design context, and engage in ideation through conversations in the conversational window.}%
    \label{fig:structure}%
    \Description{This figure illustrates the interaction between the backend and frontend of the DesignMinds system. On the left, the backend features two primary inputs: a video of a person slicing a cucumber on a cutting board, and "Design Books," which provide contextual knowledge. The video is processed by a Vision-Language Model (VLM), while the design context from the books is processed by a Large Language Model (LLM). These two models combine information to generate prompts for the frontend. In the frontend, a designer interacts with two components: a "Video Playback" window and a "Conversation Window." The video playback allows the designer to watch relevant footage, while the conversation window facilitates interaction with the system, where the designer asks questions and receives answers generated by the LLM. This setup enables designers to enhance their creative process through interaction with the system's insights.}
\end{figure}

\begin{figure}
  \centering
    \includegraphics[width=\linewidth]{ 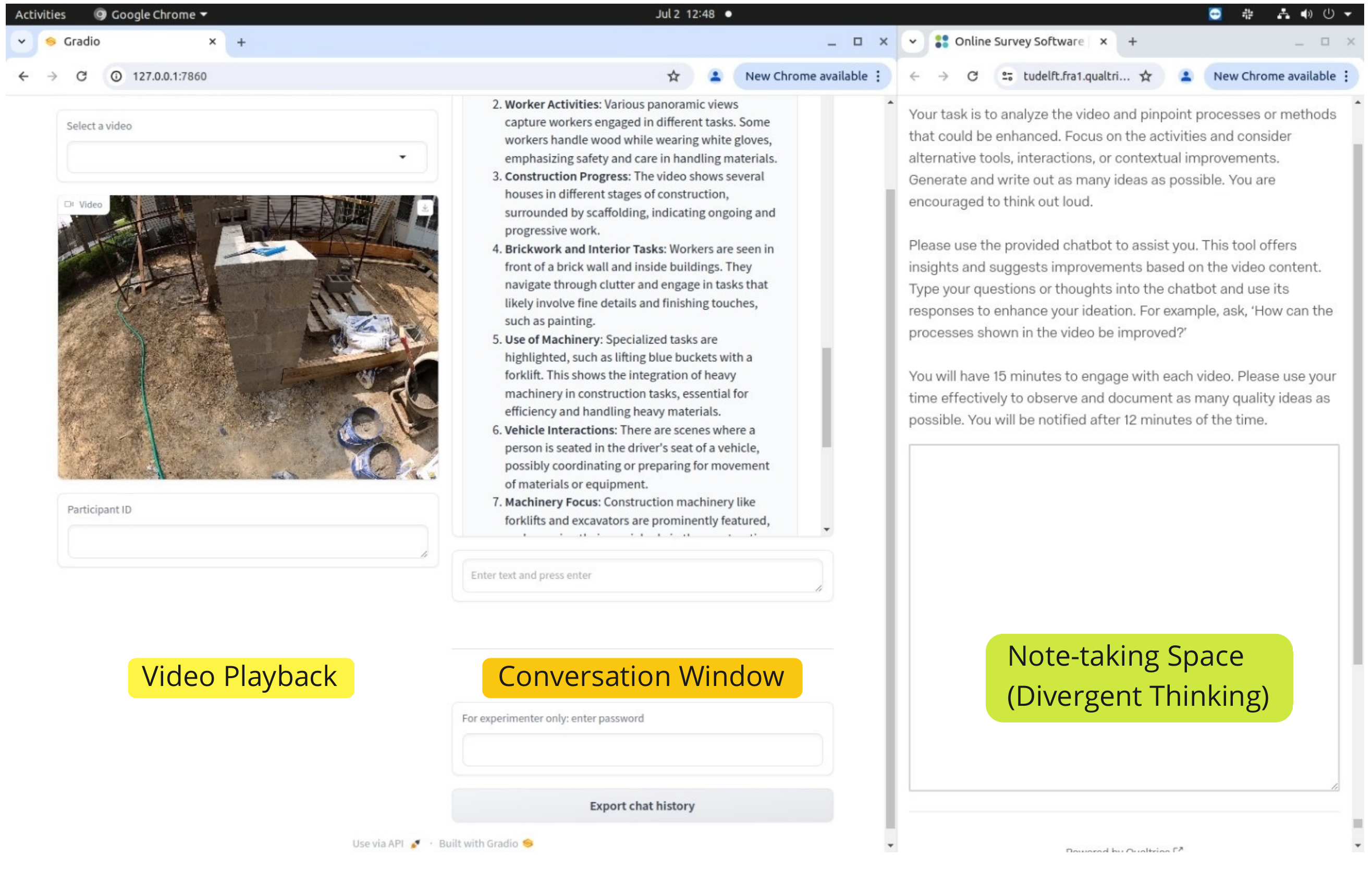}
    \caption{The interface of \proto primarily features a video player on the left and an LLM conversation window in the center. To facilitate organized ideation recording in the later study, we additionally included a note-taking space bellow a description of VBD tasks for recording participants' divergent thinking during the study tasks (see Supplementary Text \ref{instruction text}) for detailed text. When designers use \proto, the system initially performs a background pre-analysis of the video content on the left, and transitions video content to the chat interface in the center. Designers subsequently interact via chatting and generate inspiration as Divergent Thinking notes on the right.}%
    \label{fig:UI}%
    \Description{The figure shows the interface of the DesignMinds platform divided into three sections: "Video Playback" on the left, "Conversation Window" in the center, and "Note-taking Space (Divergent Thinking)" on the right. On the left, a video playback area displays a construction site with a stack of bricks, tools, and dirt. Below the video is a section for selecting videos and inputting participant IDs. In the center is the "Conversation Window," where users can input text to interact with the system, specifically designed for chatting with the LLM for insights and suggestions related to design processes. Finally, on the right, the "Note-taking Space" provides an area where users can write down their divergent thinking ideas and observations while engaging with the video content and system responses. }
\end{figure}

The development of our prototype followed the natural process of the idealization of VBD, consisting of two main parts: video comprehension and idea reflection and refinement \cite{jacob_edit_2007}. As shown in Fig. \ref{fig:structure}, \proto consists of two primary components: the backend and the front-end. The backend includes a VLM and a LLM integrated with a design knowledge repository for reference. The front-end features corresponding a video playback region alongside a conversational window. We adopted \texttt{blip2-opt-6.7b}\footnote{\url{https://huggingface.co/facebook/opt-6.7b} (last accessed: \today).}, a SOTA VLM, to interpret videos into textual descriptions. When processing a video, the VLM first extracts perceived objects from the video and utilizes built-in language connection functions to generate comprehensive textual descriptions of the entire video. These complete video descriptions then were processed by an LLM through GPT-4 API (\texttt{gpt-4-0125-preview})\footnote{\url{https://platform.openai.com/docs/models/gpt-4-turbo-and-gpt-4} (last accessed: \today).}. To generate more design-grounded suggestions, we implemented a RAG function using a text embedding model \texttt{text-embedding-ada-002}\footnote{\url{https://platform.openai.com/docs/guides/embeddings} (last accessed: \today).} on a framework of LlamaIndex\footnote{\url{https://docs.llamaindex.ai/en/stable/} (last accessed: \today).} as our \proto's professional knowledge repository for conversations. To ensure that the knowledge repository provided designer-relevant information for our LLM, we conducted a discussion on VBD literature within an independent community of designers (\textit{N} = 30). This discussion led to a vote that selected six authoritative books (1,966 pages total) with high-level methodological rigor and practical design cases for the VBD training. We then utilized the RAG function and tokenize the selected design books to feed into the knowledge repository of the LLM. We then built our front-end interface using Gradio\footnote{\url{https://www.gradio.app/} (last accessed: \today)} as illustrated in Fig. \ref{fig:UI}. The interface includes a video player and an chatbot conversation window. To test performance and enhance convenience for test users in the later study, we allocated the right portion of the screen to included a text box where users could record their ideas and inspirations. This setup allows users to review and revisit the design context using the video player, generate additional insights and ideas through the chatbot, and record their comprehensive thoughts in the text box for later use.

\section{Study}
We evaluate how our proposed \proto influences ideation in VBD tasks with a between-subject study design. Specifically, we examine whether and how the tool influences designers' effectiveness and ability to generate ideas from video content. Our assessment is structured around three key perspectives: the quality of ideas generated by designers, the cognitive processes they undergo during the ideation tasks, and their overall user experience and acceptance of the new prototype. Additionally, we analyze how designers interact with \proto from their conversation logs to better understand the ideation process. We also explore \proto' potential cognitive effects, perceived usefulness and likelihood of adoption by designers. Finally, we investigate areas for improvement and suggest ways to enhance \proto' usability and other concerns. Our study addresses the following Research Questions (RQs): 
\begin{enumerate}
  \item[RQ1] \textit{How does \proto influence the quality of ideas generated in the VBD process?} 

Divergent thinking, a concept introduced by Guilford \cite{guilford_creativity_1967, guilford_creativity_1950} acts as a foundational idea in creativity research. Divergent thinking emphasizes the generation of novel, free-flowing, and unconventional ideas, allowing for the expansion of the design space to identify innovative solutions based on available resources \cite{guilford_creativity_1950, acar_divergent_2019, baer_creativity_2014}. This approach is a key model in design ideation, where effective divergent thinking is often regarded as indicative of successful ideation \cite{jacob_design_2007}. Building on this foundation, we investigate how our \proto impacts the outcomes of divergent thinking by asking participants in two conditions (experimental group and control group) to generate creative ideas during the task. We hypothesize that \textbf{designers with AI co-ideation will exhibit higher Divergent Thinking scores compared to ideation without AI}.

\item[RQ2] \textit{How does \proto influence the way designers practice ideation in VBD?}  
    

  
Examining user behaviors is another critical aspect of evaluating the VBD ideation process, in addition to assessing the final deliverables. The behaviors exhibited during tasks reflect participants' approaches to completing the assigned tasks \cite{babiloni_mental_2019, hinson_cognitive_2003, ayaz_optical_2012}. We record their eye movements to evaluate the level of engagement and cognitive load experienced by designers in both conditions. Additionally, we conduct an in-depth analysis of the chat log history from the experimental group to understand how participants interacted with \proto. We hypothesize that \textbf{designers will experience greater engagement and, consequently, a slightly higher cognitive load in the AI-prototype-assisted condition}.
  
  \item[RQ3] \textit{What impact does \proto have on the User Experience (UX) and Technology Acceptance and Use in the VBD ideation process?} 
  
  
  The introduction of new technologies or tools to a traditional methodology can sometimes cause discomfort and decreases in UX \cite{tarafdar_impact_2007}. Understanding and evaluating technology acceptance and use also provides insights into how well users adapt to new technology, which may potentially impact the original practice. We further compare the UX and the level of acceptance and use of technology between our prototype condition and the control condition during the VBD ideation process. We hypothesize that \textbf{the newly introduced prototype will not have additional negative influence on UX and technology acceptance and use compared to traditional practices}.
\end{enumerate}

\subsection{Participants}

\begin{table}[h]
\centering
\caption{The demographics of participants' design experience, including possible responses and their values, are presented as answer frequencies (f), followed by the corresponding percentages (\%).}
\begin{tabular}{p{3.5cm} c c c}
\toprule
\textbf{Variable} & \textbf{Answer} & \textbf{f} & \textbf{\%} \\
\midrule
\multirow{3}{*}{\parbox{3.5cm}{Current design educational level}} & Bachelor & 12 & 34.29\% \\
& Master & 22 & 62.86\% \\
& PhD (ongoing) & 1 & 2.86\% \\
\midrule
\multirow{5}{*}{\parbox{3.5cm}{Experience of designing with videos (VBD)}} & Definitely not & 15 & 42.86\% \\
& Probably not & 7 & 20.00\% \\
& Might or might not & 9 & 25.71\% \\
& Probably yes & 2 & 5.71\% \\
& Definitely yes & 2 & 5.71\% \\
\midrule
\multirow{5}{*}{\parbox{3.5cm}{Experience of practicing design divergent thinking (ideation) }} & Definitely not & 2 & 42.86\% \\
& Probably not & 2 & 5.71\% \\
& Might or might not & 9 & 25.71\% \\
& Probably yes & 17 & 48.57\% \\
& Definitely yes & 5 & 14.29\% \\
\midrule
\multirow{4}{*}{\parbox{3.5cm}{Proficiency in using chatbot}}
& Never used before & 2 & 5.71\% \\
& Beginner & 7 & 20.00\% \\
& Intermediate & 13 & 37.14\% \\
& Expert & 13 & 37.14\% \\
\bottomrule
\end{tabular}
\label{tab:demographic}
\Description{The table provides demographic details of participants' design experience about educational level, experience in designing with videos (VBD), experience practicing divergent thinking (ideation), and proficiency in using chatbots. In terms of educational level, 62.86\% (22 participants) were at the Master's level, 34.29\% (12 participants) at the Bachelor's level, and 2.86\% (1 participant) were ongoing PhD students. Regarding VBD experience, 42.86\% (15 participants) reported "Definitely not," 20.00\% (7 participants) said "Probably not," 25.71\% (9 participants) indicated uncertainty ("Might or might not"), and small proportions of 5.71\% (2 participants) each selected "Probably yes" and "Definitely yes." In practicing design divergent thinking, 48.57\% (17 participants) responded "Probably yes," 14.29\% (5 participants) said "Definitely yes," while 25.71\% (9 participants) were unsure, and small proportions of 5.71\% (2 participants) each answered "Probably not" and "Definitely not." Finally, chatbot proficiency varied, with 37.14\% (13 participants) considering themselves either Intermediate or Expert, 20.00\% (7 participants) as Beginners, and 5.71\% (2 participants) having never used a chatbot.}
\end{table}
We enlisted 35 design graduates (17 females and 18 males) from the design faculty at our university, following approval from the ethics board and confirming that none had any cognitive impairments. The participants, who are either university students (BSc \& MSc) or PhD candidates, had an average age of 25.4 years (SD = 2.31) and an average of 2.4 years of design experience (SD = 1.14). Table \ref{tab:demographic} presents the demographics of participants involved in the study, including their educational levels, self-assessed familiarity with VBD experience and ideation, as well as their proficiency in using chatbots like ChatGPT. In addition, participants with visual acuity below 20/20 were instructed to wear contact lenses before participating. All participants were fully informed and provided consent before the experiment began.

\subsection{Apparatus}
In our experiment, we evaluated our system in an office setting with consistent lighting. The system was set up to operate as localhost on a desktop computer within the lab environment. Fig. \ref{fig:studyScence} illustrates the lab setup where participants engaged with the system. Alongside standard office equipment such as a keyboard, mouse, and speaker, participants were asked to wear eye-tracking glasses (NEON type from Pupil Labs\footnote{\url{https://pupil-labs.com/products/neon/} (last accessed: \today).}, sampling rate 200 Hz). These glasses were connected to an Android phone via a USB-C cable to record eye-movement data. The data collected included pupil dilation changes, gaze positions, and blink patterns. The monitor was positioned in a comfortable visual range of 55 centimeters from the user-facing edge of the table and was a 22-inch screen tilted 15 degrees below the participants' horizontal line of sight \cite{monitor_position_2023}. Additionally, we placed four AprilTags\footnote{\url{https://april.eecs.umich.edu/software/apriltag} (last accessed: \today).} on each corner of the screen (see Fig. \ref{fig:studyScence}) to allow the eye-tracking glasses to detect the screen’s surface accurately and define the \proto' interface as the area of interest (AOI).
\subsection{Measures}
\subsubsection{Subjective Measures}
\begin{itemize}
\item \textbf{Evaluation of Divergent Thinking (RQ1)}: we employ the concept of divergent thinking outlined by Guilford \cite{guilford_creativity_1967, guilford_creativity_1950}, and assess it through the following three dimensions:

\subitem \textbf{- Fluency}: a measurement captures the quantity of comprehensive ideas generated. Each idea must be sufficiently detailed in terms of purpose and functionality to be clearly understood.

\subitem \textbf{- Flexibility}: a measurement  evaluates the range of different domains and subdomains covered by the ideas, reflecting the diversity of the ideation process.

\subitem \textbf{- Originality}: a measurement evaluates the uniqueness of ideas, measured by their statistical infrequency, and is evaluated using a 7-point Likert scale.

\item \textbf{Chat Log history (RQ2)}: the intermediate conversation history made by participants in the experimental group with AI co-ideation with the chatbot portion of \proto. 

\item \textbf{Unified theory of acceptance and use of technology (UTAUT) (RQ3)}: a widely recognized model for assessing how users accept and adopt information technology considers the perceived likelihood of adoption. This likelihood is influenced by five key constructs: performance expectancy, effort expectancy, attitude toward using technology, anxiety, and behavioral intention to use the system \cite{venkatesh_user_2003, williams_unified_2015, attuquayefio_using_2014}. 

\item \textbf{User Experience Questionnaire (UEQ) (RQ3)}\footnote{\url{https://www.ueq-online.org/} (last accessed: \today). }:
a questionnaire designed to measure UX in interactive products uses a benchmarking method that organizes raw UEQ scores into categories such as efficiency, perspicuity, dependability, originality, and stimulation \cite{schrepp_applying_2014, schrepp_construction_2017}.

\end{itemize}
\subsubsection{Objective Measures (RQ2)}
\begin{itemize}
\item \textbf{Pupil Dilation}: an involuntary physiological response where the pupils widen during assigned tasks.
\item \textbf{Fixation Rate and Duration}: measurements describe how often and how long the eyes remain stationary on a specific point during tasks, with fixation rate indicating the frequency of these pauses and fixation duration indicating the length of time the eyes stay still in one position.
\item \textbf{Blink Rate and Duration}: measurements describe how often and how long the eyelids rapidly closing and opening during tasks. 
\item \textbf{Saccade Rate and Speed}: measurements describe how often and how quickly the eyes perform fast and conjugate movements from one fixed point to another.
\end{itemize}

\subsection{Procedure}
\begin{figure}
  \centering
    \includegraphics[width=.6\linewidth]{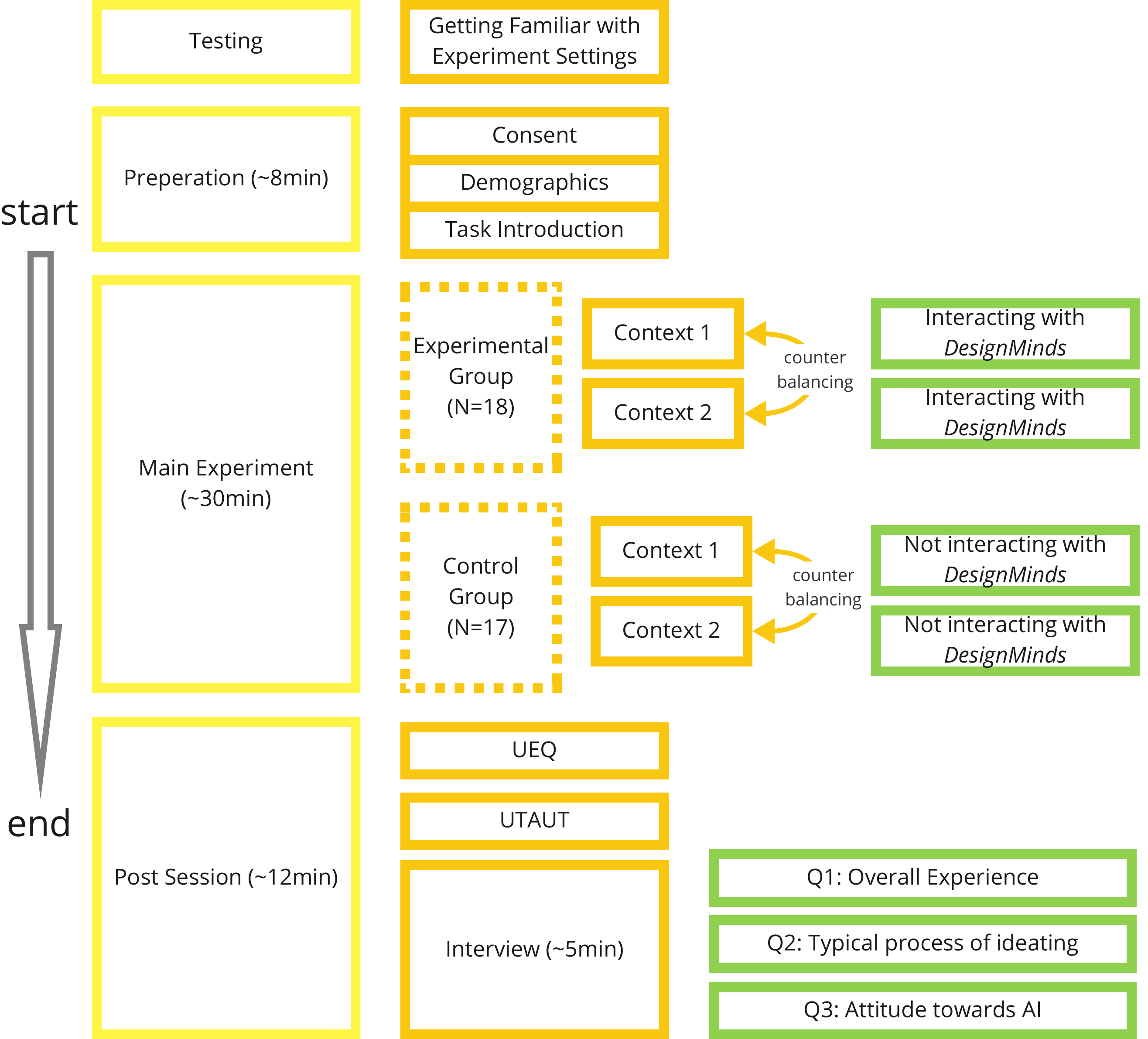}
    \caption{During the study, participants were initially asked to familiarize themselves with both the environment and \proto (Testing). They received instructions on the components of the prototype and how to interact with it. Following this, participants completed consent and demographic forms for background information. They were then provided with instructions for the tasks (Preparation). Participants were randomly divided into two groups: the experimental group, which interacted with the chatbot \proto, and the control group, where participants continued their usual practice for design inspiration. Each participant group was assigned two tasks with different design contexts, presented in a counterbalanced order. In the next session (Post Session), participants were asked to complete the UEQ  and UTAUT questionnaires. Finally, they were interviewed for about 5 minutes on three topics: overall experience, typical ideation process, and their attitudes towards AI.}%
    \label{fig:studyPrecedure}%
    \Description{The figure shows the flow of an experimental design divided into four stages. The first stage is "Testing," where participants are asked to "Get Familiar with Experiment Settings." The second stage is "Preparation" (approximately 8 minutes), which includes tasks: consent, demographics, and task introduction. The third stage is the "Main Experiment" (approximately 30 minutes), where participants are divided into two groups: the "Experimental Group" (N=18) and the "Control Group" (N=17). Both groups engage with two contexts (Context 1 and Context 2) with counterbalancing. The experimental group interacts with DesignMinds, while the control group does not. The final stage is the "Post Session" (approximately 12 minutes), where participants complete two questionnaires, including UEQ and UTAUT, followed by a short interview (approximately 5 minutes). The figure also outlines specific questions explored in the study, such as Q1 (Overall Experience), Q2 (Typical process of ideating), and Q3 (Attitude towards AI).}
\end{figure}
\subsubsection{Preparation and Main Experiment Session}
Before the study began, participants were assigned to either the control or experimental group using de-identified IDs. Participants were individually invited to the lab according to their scheduled times and took their designated positions in front of the monitor (see Fig. \ref{fig:studyScence}). With assistance, they adjusted the monitor's height and tilt angle based on their measured height and seating posture. They were then introduced to the study apparatus, including the user interface (shown in Fig. \ref{fig:UI}) relevant to their assigned group, how to wear the eye-tracking glasses, and briefed on the study procedure. After this introduction, participants were asked to complete a consent form and provide demographic information, including their experience with design ideation from videos, general design experience, and familiarity with using chatbots. Then they received instructions (see Supplementary Text \ref{instruction text}) on the tasks they were required to complete.

Following the preparatory phase, participants in each group were shown two video tasks depicting contexts of cooking and construction, with the order of presentation counterbalanced. These videos were sourced from Ego4D\footnote{\url{https://ego4d-data.org/} (last accessed: \today). }, a large-scale video dataset frequently employed for benchmark and HCI research \cite{grauman_ego4d_2021}. Each video was approximately 3 minutes in length to ensure brevity, considering the total maximum testing time of 15 minutes per video. In the experimental group, participants were instructed to use the defined UI shown in Fig. \ref{fig:UI} to watch video playback, interact with the chatbot, and make notes in the designated note-taking space to record divergent thinking. In contrast, for the control group, the chatbot was hidden, and participants were asked to proceed with design ideation on the note-taking space from the videos as they normally would. Participants were notified at the 12-minute mark of each task that they had 3 minutes remaining. This alert was designed to keep them informed of the time constraints and allow them to prepare for the conclusion of the current task. This process was repeated for both videos.

\subsubsection{Post Session}
Upon completion, the eye-tracking recordings were halted. Participants were then asked to evaluate their experience using UEQ and UTAUT questionnaires. Following the questionnaires, a brief interview of approximately 5 minutes was conducted. Participants were asked three main topics: their overall experience during the two video tasks, their performance in the ideation process, and their attitudes toward AI after having the experiment.

\section{Results}
\subsection{Divergent Thinking Analysis (RQ1)}
\begin{figure}[h]
  \centering
    \includegraphics[width=.55\linewidth]{ 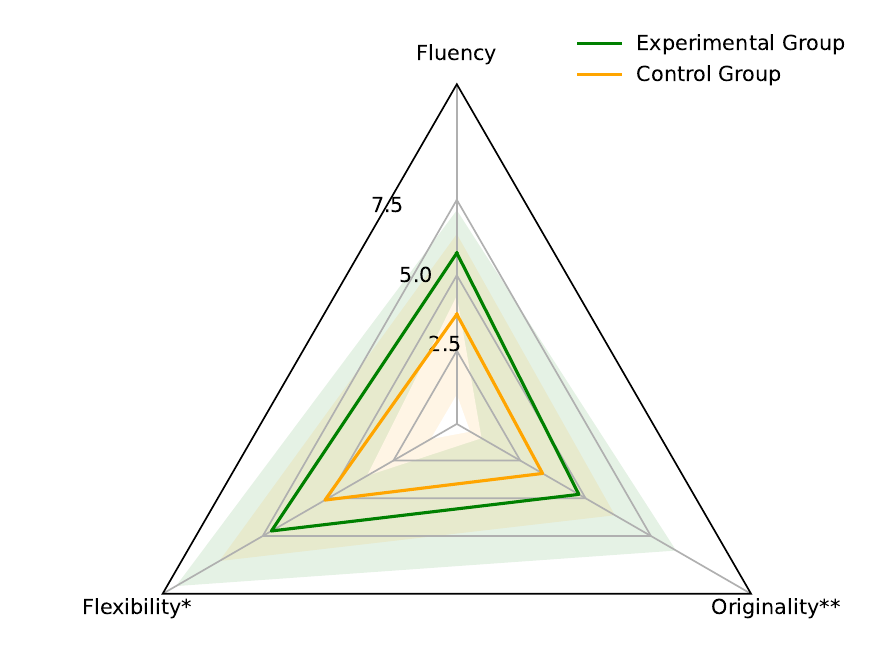}
    \caption{Radar chart depicting the evaluation scores of design thinking across raters for the experimental and control groups. Errors are indicated by shaded regions. Attributes marked with asterisks (* or **) represent significant differences. * denotes 0.01 < p < .05, and ** denotes p < .001.}%
    \label{fig:divergentThinking}%
    \Description{The radar chart compares the performance of two groups—Experimental Group (in green) and Control Group (in orange)—across three variables: "Fluency," "Flexibility," and "Originality." Each axis represents one of these variables, with a scale from 0 to 7.5. The chart shows that the experimental group consistently outperforms the control group across all three categories, as seen by the larger coverage area of the green line compared to the orange line. The experimental group achieves notably higher scores in "Originality," while the control group lags in all variables. The detailed data for each variable is as follows: Fluency shows a mean of 5.73 (standard deviation: 4.32) for the Experimental Group, and a mean of 3.71 (standard deviation: 1.05) for the Control Group. Flexibility shows a mean of 7.17 (standard deviation: 3.51) for the Experimental Group, and a mean of 5.12 (standard deviation: 1.07) for the Control Group. Originality shows a mean of 4.74 (standard deviation: 1.02) for the Experimental Group, and a mean of 3.35 (standard deviation: 0.58) for the Control Group. A single asterisk sign (*) is marked on "Flexibility," while a double asterisk sign (**) is marked on "Originality," indicating significant differences in these variables.}
\end{figure}
To address the quality of ideas generated in the VBD process as proposed in RQ1, we collected the divergent thinking texts from both groups. We then recruited three independent raters to evaluate the ideation results based on fluency, flexibility, and originality, using a predetermined set of criteria (See supplementary text \ref{divergent thinking criteria}) \cite{guilford_creativity_1950}. We then performed a quantitative analysis of the rating scores for both the experimental and control groups. As shown in Fig. \ref{fig:divergentThinking}, we observed a significant main effect on the average ratings for flexibility and originality (independent t-test \(t(33)t=2.304\), \(p = .014\); \(t(33)t=4.674\), \(p < .001\)). \textbf{The average scores for both flexibility} (7.17 ± 3.511 points) \textbf{and originality} (4.74 ± 1.018 points) \textbf{in the experimental group were significantly higher than those in the control group} (flexibility: 5.12 ± 1.074 points; originality: 3.35 ± 0.583 points). However, there was no significant main effect on the rating for fluency between the two groups (independent t-test \(t(33)t=1.885\), \(p = .068\)). Additionally, Krippendorff’s Alpha was calculated to assess the internal consistency of the three raters' judgments on the categories of divergent thinking. We observed a moderate agreement among the raters, with an average Krippendorff’s Alpha of $\alpha$ = .702 (95\% CI, .245 to 1), p < .001.
\subsection{Design Ideation Process (RQ2)}
\subsubsection{Eye-tracking measures} 
\begin{figure}
  \centering
  \begin{subfigure}{.4\linewidth}
    \centering
    \includegraphics[width=\linewidth]{ 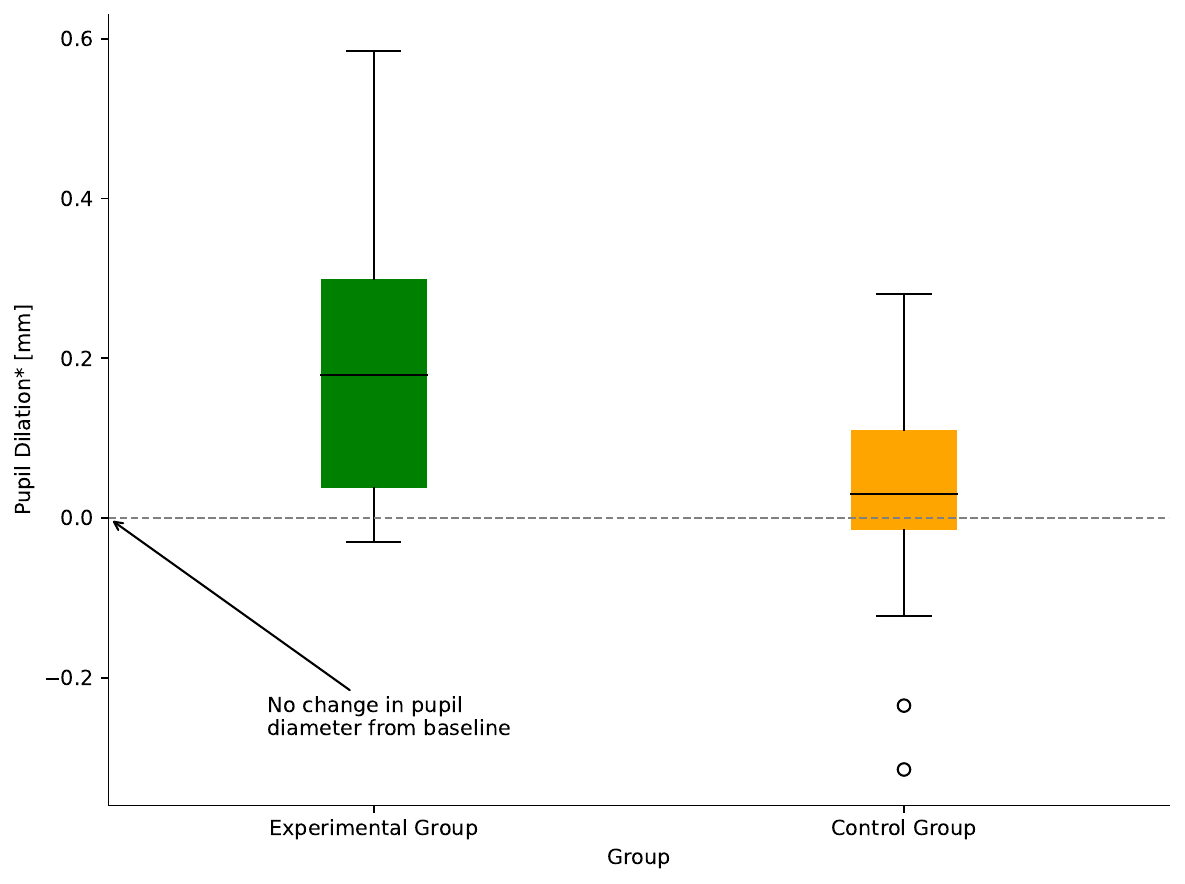}
    \caption{Participants in the experimental group exhibited significantly greater pupil dilation compared to the control group. The dashed line at 0 millimeter on y axis represents no change in pupil diameter relative to the baseline, when participants were not engaged in ideation tasks.}
    \label{fig:pupil}
    \Description{The box plot (a) compares pupil dilation in millimeters between two groups: "Experimental Group" shown in green and "Control Group" in orange. The y-axis ranges from -0.2 mm to 0.6 mm, and a dashed horizontal line at 0 mm indicates no change from the baseline pupil diameter. The experimental group has a mean dilation of 0.18 mm and a standard deviation of 0.21 mm, shows a median significantly above the baseline with the interquartile range mostly above zero. Some outliers extend above the upper whisker. The control group has a mean dilation of 0.03 mm and a standard deviation of 0.15 mm with a median near zero. The control group also has smaller variations in pupil dilation with a couple of outliers below the lower whisker.}
  \end{subfigure}
\hspace{1em} 
  \begin{subfigure}{.48\linewidth}
    \centering
    \includegraphics[width=\linewidth]{ 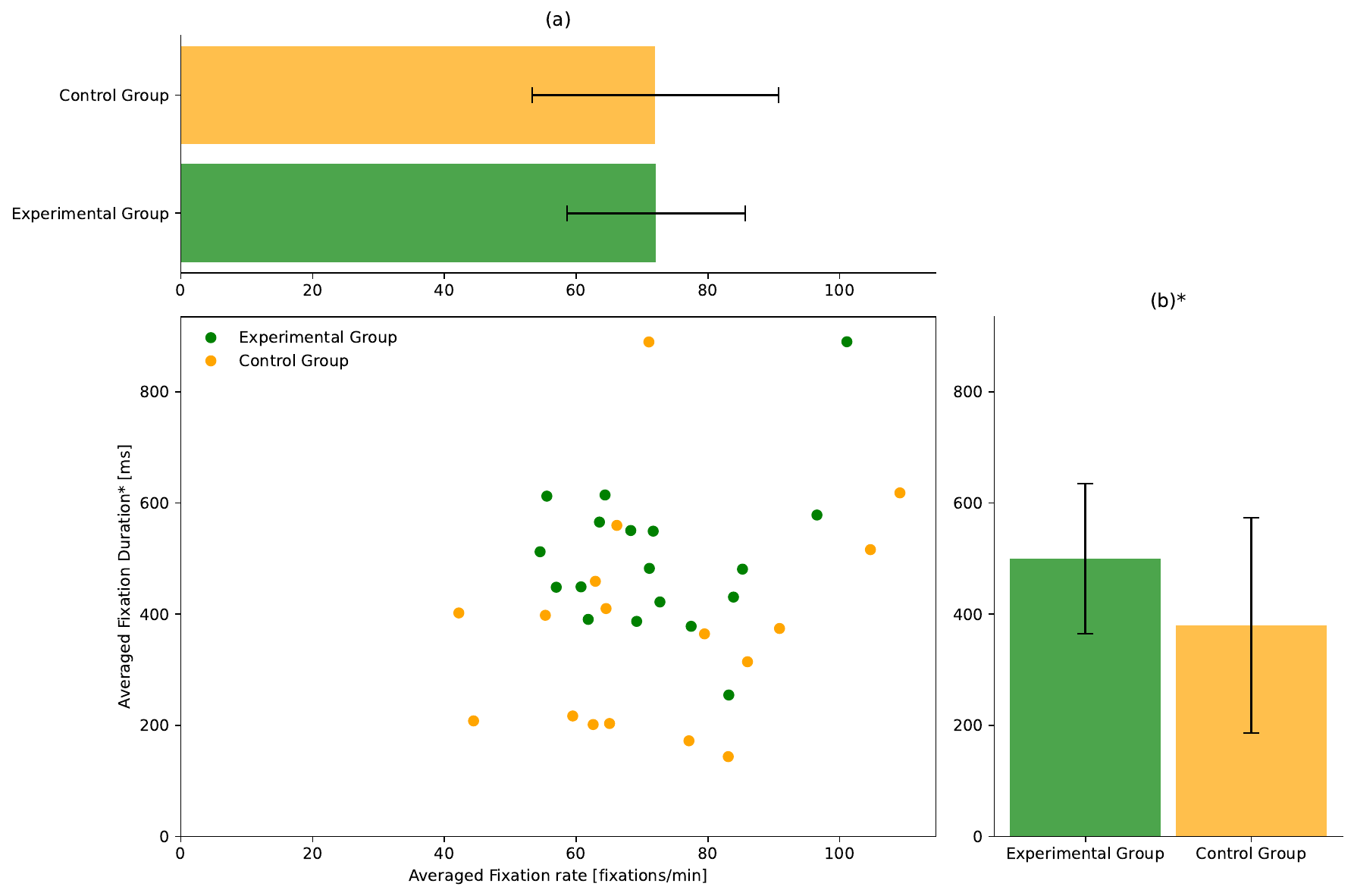}
    \caption{In subplot (b-a), no significant difference was observed in the averaged fixation rate between the groups. In subplot (b-b) indicated by an asterisks(*), participants in the experimental group exhibited a significantly higher fixation duration compared by control group.}%
    \label{fig:fixation}%
    \Description{The subplot (b) contains two subplots related to eye fixation movement in experimental and control groups. Subplot (b-a) combines a scatter plot with a bar graph of the averaged fixation rate per minute, showing no significant difference between the groups, with the experimental group having a mean fixation rate of 72.1193 (SD: 13.52) and the control group at 72.02 (SD: 18.71). Subplot (b-b) combines a scatter plot with a bar graph comparing average fixation durations. The experimental group exhibits a significantly higher mean fixation duration of 499.93 milliseconds (SD: 135.05) compared to the control group's 379.62 milliseconds (SD: 193.37), with an asterisk (*) indicating statistical significance.}
  \end{subfigure}
  \hspace{1em} 
  \begin{subfigure}{.48\linewidth}
    \centering
    \includegraphics[width=\linewidth]{ 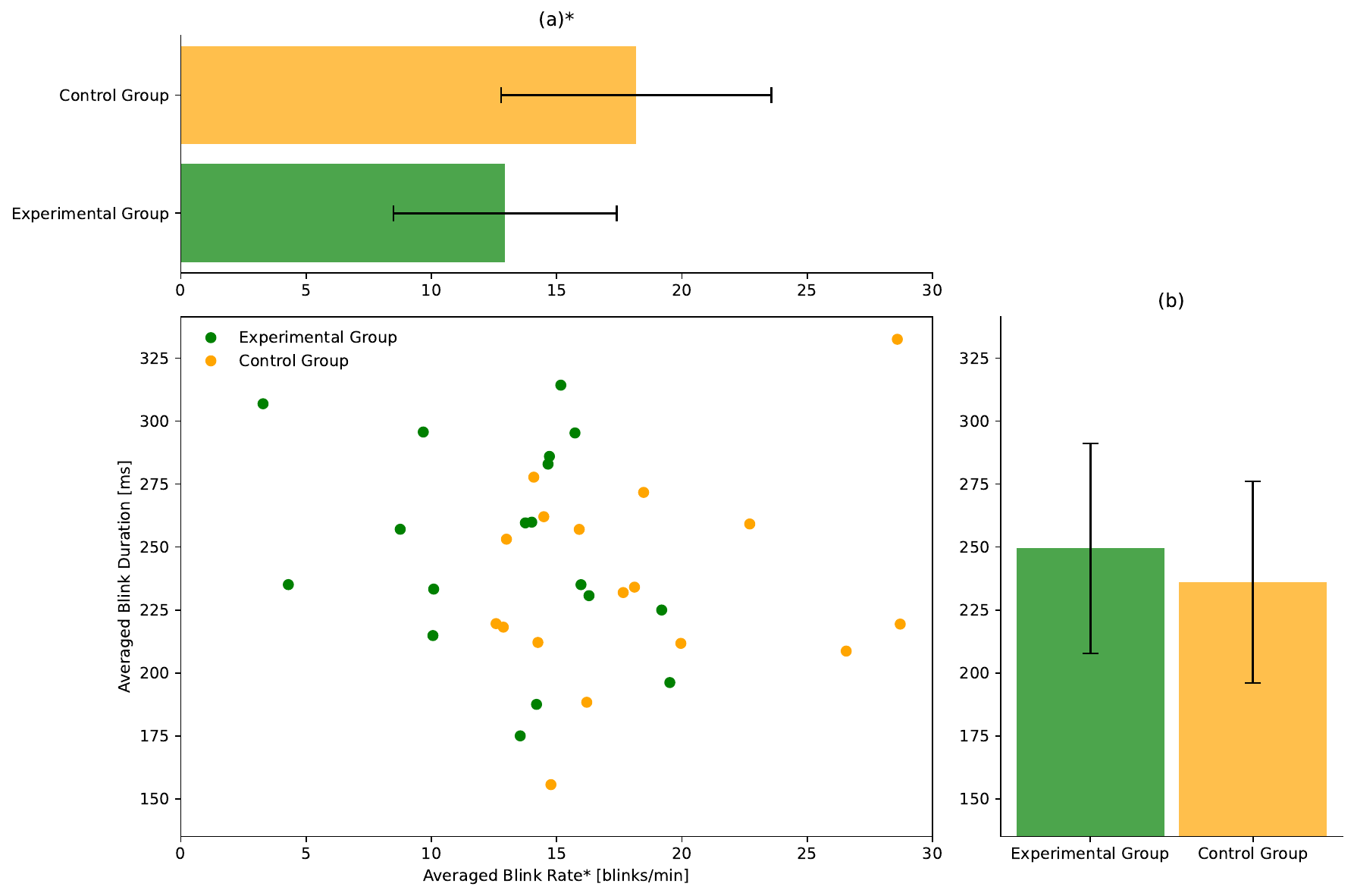}
    \caption{In subplot (c-a), participants in the experimental group exhibited a significantly lower blink rate, as indicated by an asterisks(*). In contrast, (c-b) shows no significant difference was observed in blink duration between the groups.}%
    \label{fig:blink}%
    \Description{The subplot (6c) contains two subsequent subplots, comparing blink rate and blink duration between the experimental group and the control group. The scatter plots in both subplots depict the relationship between the averaged blink rate (blinks per minute) and the averaged blink duration (milliseconds) for individual participants in the experimental group (green dots) and the control group (orange dots). Subplot (c-a) shows a bar graph comparing the average blink rate between the two groups. The experimental group has an average blink rate of 12.94 blinks/min (SD: 4.46), while the control group has a significantly higher average blink rate of 18.17 blinks/min (SD: 5.40), indicated by an asterisk (*) for statistical significance. Subplot (c-b) displays a bar graph comparing the average blink duration for the two groups. The experimental group has a slightly longer average blink duration of 249.46 milliseconds (SD: 41.67) compared to the control group's 236.06 milliseconds (SD: 40.16).}
  \end{subfigure}
  \vspace{1em} 
  \begin{subfigure}{.48\linewidth}
    \centering
    \includegraphics[width=\linewidth]{ 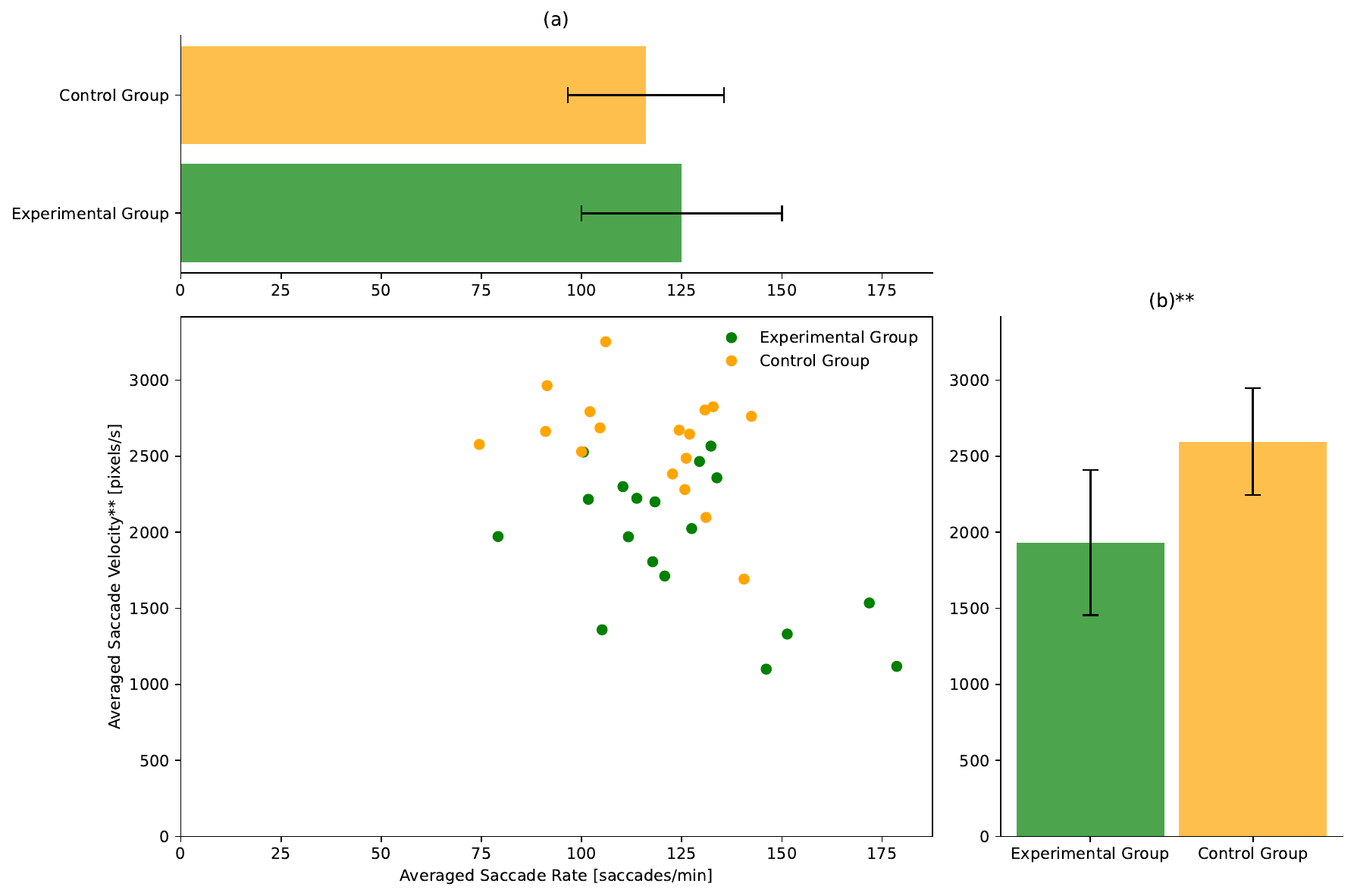}
    \caption{In subplot (d-a), no significant difference found compared averaged saccade rate across the two groups. In subplot (d-b) with asterisks(**), participants in the experimental group exhibited a significantly lower number of average velocity in saccade.}%
    \label{fig:saccade}%
    \Description{The subplot (6d) contains two subsequent subplots (d-a and d-b) comparing saccade rate and saccade velocity between an experimental group and a control group. Subplot (d-a) shows a bar graph comparing the average saccade rate (saccades per minute) for both groups. The experimental group has an average saccade rate of 124.98 saccades/min (SD: 25.01), slightly higher than the control group's average of 116.05 saccades/min (SD: 19.51), but the difference is not statistically significant. Subplot (d-b) shows a bar graph comparing the average saccade velocity (in pixels/second). The experimental group has an average saccade velocity of 1933.18 pixels/sec (SD: 477.33), while the control group has a significantly higher average saccade velocity of 2595.63 pixels/sec (SD: 351.45), indicated by two asterisks (**) to denote a high statistical significance (p<.001).}
  \end{subfigure}

  \caption{Plots displaying the average pupil dilation (\ref{fig:pupil}), fixation rate and duration (\ref{fig:fixation}), average blink rate and duration (\ref{fig:blink}), and average saccade rate and velocity (\ref{fig:saccade}) for the experimental and control groups. Accompanying histograms with error bars are also provided for each measure. Attributes and subplots marked with asterisks (* or **) represent significant differences. * denotes 0.01 < p < .05, and ** denotes p < .001.}
  \label{fig:eyetracking}

\end{figure}

We first analyzed the eye-tracking results from both groups. As shown in Fig. \ref{fig:blink}, a significant main effect was observed in the average pupil dilation between the experimental and control groups (independent t-test \(t(33)t=2.933\), \(p = .021\)). The dashed line in the subplot represents 0 millimeters which indicates there was no change from participant's baseline pupil diameter during non-tasked time. Compared to the baseline, \textbf{participants in the experimental group exhibited an average dilation of 0.15 mm more than those in the control group during the ideation task} (experimental \(std = 0.206\); control \(std = 0.152\)). We then examined the gaze fixation rate per minute and the average fixation duration across the two groups. As shown in Fig. \ref{fig:fixation}, no significant main effect (independent t-test \(t(33)t=0.795\), \(p = .986\)) was observed in the average fixation rate (see subplot (a)). Interestingly, \textbf{participants in the experimental group exhibited an average fixation duration that was significantly 120.31 milliseconds} (\(std = 135.053\)) \textbf{longer than that of the control group within the AOI } (\(std = 193.366\); independent t-test \(t(33)t=1.567\), \(p = .039\)). Additionally, as shown in subplot (a) of Fig. \ref{fig:blink}, we observed a significant main effect in the average blink rate (independent t-test \(t(33)t=0.557\), \(p = .004\)). \textbf{Participants in the experimental group on average blinked 5.23} (experimental \(std = 4.459\); control \(std = 5.400\)) \textbf{less times per minute than those in the control group}. However, no significant difference was found in the average blink duration between the two groups (independent t-test \(t(33)t=0.226\), \(p = .340\)). No significant main effect was observed in the average saccade rate between the two groups shown in Fig. \ref{fig:saccade}(independent t-test \(t(33)t=0.252\), \(p = .249\)). However, there was a significant increase in saccade velocity in the experimental group compared to the control group (independent t-test \(t(33)t=3.171\), \(p < .001\)). On average, \textbf{participants in the experimental group performed 662.45 pixels per second  faster saccades than those in the control group within the AOI} (experimental \(std = 477.332\); control \(std = 351.452\)).

\subsubsection{Chat log analysis}
\label{sec:chatlog}In addition to eye-tracking measurements, we conducted an in-depth analysis of the conversation logs from the experimental group. We utilized both qualitative and quantitative methods to better understand what occurred during the augmented design ideation processes with \proto. We categorized the questions that participants asked as follows:

\begin{itemize}
\item [(a)] Questions about design opportunities (N=16): The majority of questions posed by participants (P1-3, P5, P7-17, and P19) focused on suggestions or ideas for improving the processes depicted in the videos. These inquiries typically emerged after participants had gained an understanding of the video's content and identified key areas of interest for potential design opportunities. For instance, some designers, such as P2 and P8, sought initial inspiration to begin their designs by asking, "How can the processes shown in the video be improved?" (P2) and "What can be improved?" (P8). Others (P3, P9, P12, and P19) aimed to build upon existing ideas and leveraged the LLM to further extend their concepts. These participants asked questions such as, "What do you suggest to avoid using hands directly when handling food during cooking?" (P9), "Can you recommend structures that allow a construction worker to lift heavy objects without carrying them?" (P12), and "What are the consequences of not using fitted kitchen tools for the task?" (P19).

\item[(b)] General video content understanding (N=13): Many participants (P2-4, P6-8, P10, P11, P13, P15, P16, P17 and P19) utilized the video comprehension capabilities of \proto to gain a comprehensive understanding of the content presented in the videos. Participants frequently inquired about the events occurring in the video or sought clarification on specific actions or objects they found unclear. Some participants employed a \proto-first strategy, initiating their ideation processes by querying the LLM about the video's content. For example, common inquiries included, "What is this video about?" (P2), "List the steps of the activities." (P6), "What dish is he making?" (P11), and "Can you tell me what's happening in the video?" (P15). Others used \proto to validate their observations, asking questions such as, "Are they cutting the edge in a straighter line?" (P10) and "This video was about how to cut an avocado, right?" (P17). Additionally, a subset of participants posed higher-level, reflective questions about the video's content, such as P19, who asked, "What is the goal of what they are doing during the construction work?"

\item[(c)] Understanding and Ideation from Specific Scene Settings (N=10): A subset of participants (P3, P6, P7, P9, P12-14, P16, P17 and P19) sought to utilize \proto to gain a deeper understanding of specific scene settings depicted in the videos. Unlike the broader inquiries in category (b), these participants focused on more narrowly defined actions within a given context. For example, when viewing a scene where an individual attempts to retrieve food from a sealed jar, P6 asked the LLM, "What are some ways to lock a jar automatically?" Similarly, P9 used the prototype as a tool for identifying specific items, asking, "What is the tool called that slices cheese in this video?" P14 inquired about strategies for organizing kitchen utensils, asking, "Can you combine the relocation ideas for kitchen tools?" In the context of construction, P16 sought detailed advice by asking, "How can I make sure that the men operate heavy machinery safely?" while P17 questioned, "Which is more efficient: adding an extra step in the process or using two different tools?"

\item[(d)] Combining Impressions with Opinion-Based Queries (N=4): Some participants (P6, P11, P16, P19) went a step further by integrating their own impressions with their questions and aksed for pinion-based suggestions. For instance, P11, while observing a scene involving three workers in a construction setting, asked, "Don't you think the space is crowded for 3 people?" The participant here showcased a critical evaluation of the scene. Similarly, other participants framed their questions in a way that encouraged critical thinking. For example, P19 asked, "What happens if you don't use fitted kitchen tools for the job?" 
\end{itemize}
Additionally, we conducted correlation tests to explore the relationship between traits from the chat logs during ideation and the quality of the final ideation outcomes, measured by three attributes: fluency, flexibility, and originality (see Fig. \ref{fig:divergentThinking}). We analyzed the conversation history and computed the average number of chat turns participants made with the prototype, the average number of words in each question asked and response generated, and the number of follow-up ideas generated for each participant in the experimental group. As shown in Table \ref{tab:chat log coorelation}, Pearson product-moment correlation tests were conducted to measure the relationship between chat log variables and ideation quality. There was \textbf{a strong, positive correlation between the average number of words in each participant's question and the originality of the ideas ultimately generated}, which was statistically significant ($\rho$ = .500, \(n\) = 18, \(p\) = .034). Similarly, \textbf{a strong and significantly positive correlation was found between the average number of words in each generated answer and the fluency} ($\rho$ = .636, \(n\) = 18, \(p\) = .005), \textbf{flexibility} ($\rho$ = .743, \(n\) = 18, \(p\) < .001), \textbf{and originality} ($\rho$ = .652, \(n\) = 18, \(p\) = .003) of the ideation quality. In addition, \textbf{a strong and significantly positive correlation was also observed between the average number of ideas generated from the prototype and both the fluency and flexibility} of the ideation quality ($\rho$ = .749, \(n\) = 18, \(p\) < .001; $\rho$ = .782, \(n\) = 18, \(p\) < .001).

\begin{table}[h]
\centering
\caption{Table of Pearson's correlation coefficients ($\rho$) and their p-values for four test variables from the analysis of the intermediate chat log and three ideation quality variables (see Fig. \ref{fig:divergentThinking}). Significant correlations are indicated by ** or * based on the p-values (see notes).}
\begin{tabular}{p{3.5cm} c c c}
\toprule
\multirow{2}{*}{\textbf{Chat Log Variable}} & \textbf{Ideation} & \textbf{Pearson's correlation} & \multirow{2}{*}{\textbf{\textit{P}-value}} \\
& \textbf{Quality} & \textbf{coefficient} & \\
\midrule
\multirow{3}{*}{\parbox{3.5cm}{Avg. Nr. of Chat Turns}} & Fluency & 0.261 & 0.296 \\
& Flexibility & 0.126 & 0.619 \\
& Originality & -0.313 & 0.206 \\
\midrule
\multirow{3}{*}{\parbox{3.5cm}{Avg. Nr. of Words in Each Question Asked}}
 & Fluency & 0.218 & 0.385 \\
& Flexibility & 0.318 & 0.198 \\
& Originality & .500* & 0.034 \\
\midrule
\multirow{3}{*}{\parbox{3.5cm}{Avg. Nr. of Words in Each Answer Generated}} & Fluency & .636** & 0.005 \\
& Flexibility & .743** & <.001 \\
& Originality & .652** & 0.003 \\
\midrule
\multirow{3}{*}{\parbox{3.5cm}{Avg. Nr. of Ideas Generated}} & Fluency & .794** & <.001 \\
& Flexibility & .782** & <.001 \\
& Originality & 0.398 & 0.102 \\
\bottomrule
\end{tabular}
\\ \small \textbf{Notes}: Pearson's correlation test (two-tailed) is significant at the **p < 0.01 and *p < 0.05.
\label{tab:chat log coorelation}
\Description{The table shows the Pearson's correlation coefficients and p-values for four chat log variables analyzed against three ideation quality metrics: Fluency, Flexibility, and Originality. For the average number of chat turns, the correlation with Fluency is 0.261 (p = 0.296), the correlation with Flexibility is 0.126 (p = 0.619), and the correlation with Originality is -0.313 (p = 0.206). For the average number of words in each question asked, the correlation with Fluency is 0.218 (p = 0.385), the correlation with Flexibility is 0.318 (p = 0.198), and the correlation with Originality is 0.500* (p = 0.034). For the average number of words in each answer generated, the correlation with Fluency is 0.636** (p = 0.005), the correlation with Flexibility is 0.743** (p < 0.001), and the correlation with Originality is 0.652** (p = 0.003). For the average number of ideas generated, the correlation with Fluency is 0.794** (p < 0.001), the correlation with Flexibility is 0.782** (p < 0.001), and the correlation with Originality is 0.398 (p = 0.102). A follow-up note indicates "Pearson’s correlation test (two-tailed) is significant at the **p < 0.01 and *p < 0.05".}
\end{table}

\subsection{UX, Technology Acceptance and Use (RQ3)}
\label{sec:uxandtam}
\begin{figure}
  \centering 
  \begin{subfigure}{.8\linewidth}
    \centering
    \includegraphics[width=\linewidth]{ 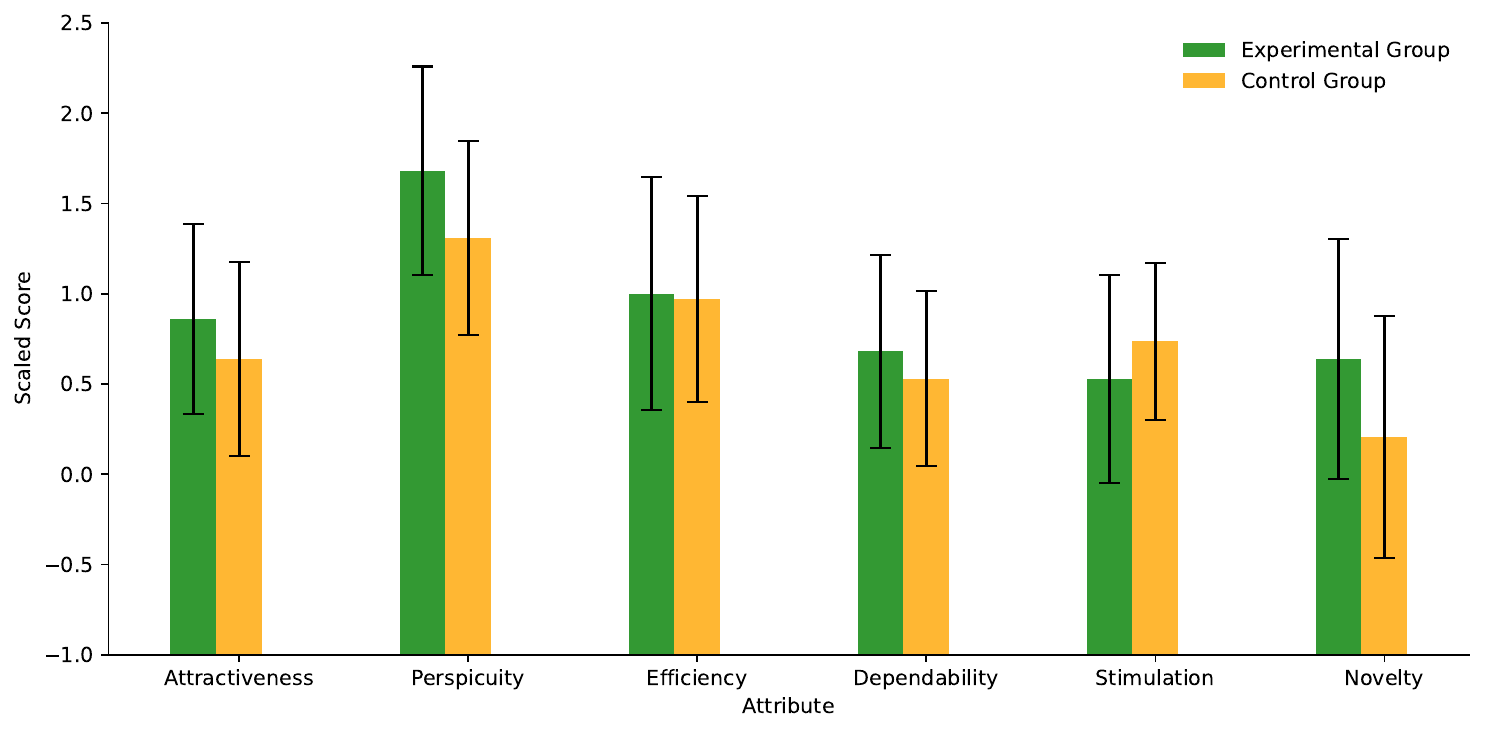}
    \caption{Scaled average values for measuring UX (UEQ) which include attributes—Attractiveness, Perspicuity, Efficiency, Dependability, Stimulation, and Novelty—compared between the experimental and control groups.}%
    \label{fig:UEQ}%
    \Description{The bar graph compares scaled average values for six user experience (UX) attributes—Attractiveness, Perspicuity, Efficiency, Dependability, Stimulation, and Novelty—between the experimental and control groups. For Attractiveness, the experimental group scored slightly higher with a mean of 0.86 (SD: 0.52) compared to the control group's 0.64 (SD: 0.54). Perspicuity also favored the experimental group with a mean score of 1.68 (SD: 0.58) versus 1.31 (SD: 0.54) for the control group. Efficiency was nearly identical between the two, with the experimental group at 1.00 (SD: 0.65) and the control group at 0.97 (SD: 0.57). For Dependability, the experimental group again had a higher mean score of 0.68 (SD: 0.54) compared to 0.53 (SD: 0.49) for the control group. However, in Stimulation, the control group scored higher with 0.74 (SD: 0.43) compared to the experimental group's 0.53 (SD: 0.58). Lastly, for Novelty, the experimental group scored higher at 0.64 (SD: 0.66) compared to the control group's 0.21 (SD: 0.67).}
  \end{subfigure}
  \vspace{1em} 
  \begin{subfigure}{.8\linewidth}
    \centering
    \includegraphics[width=\linewidth]{ 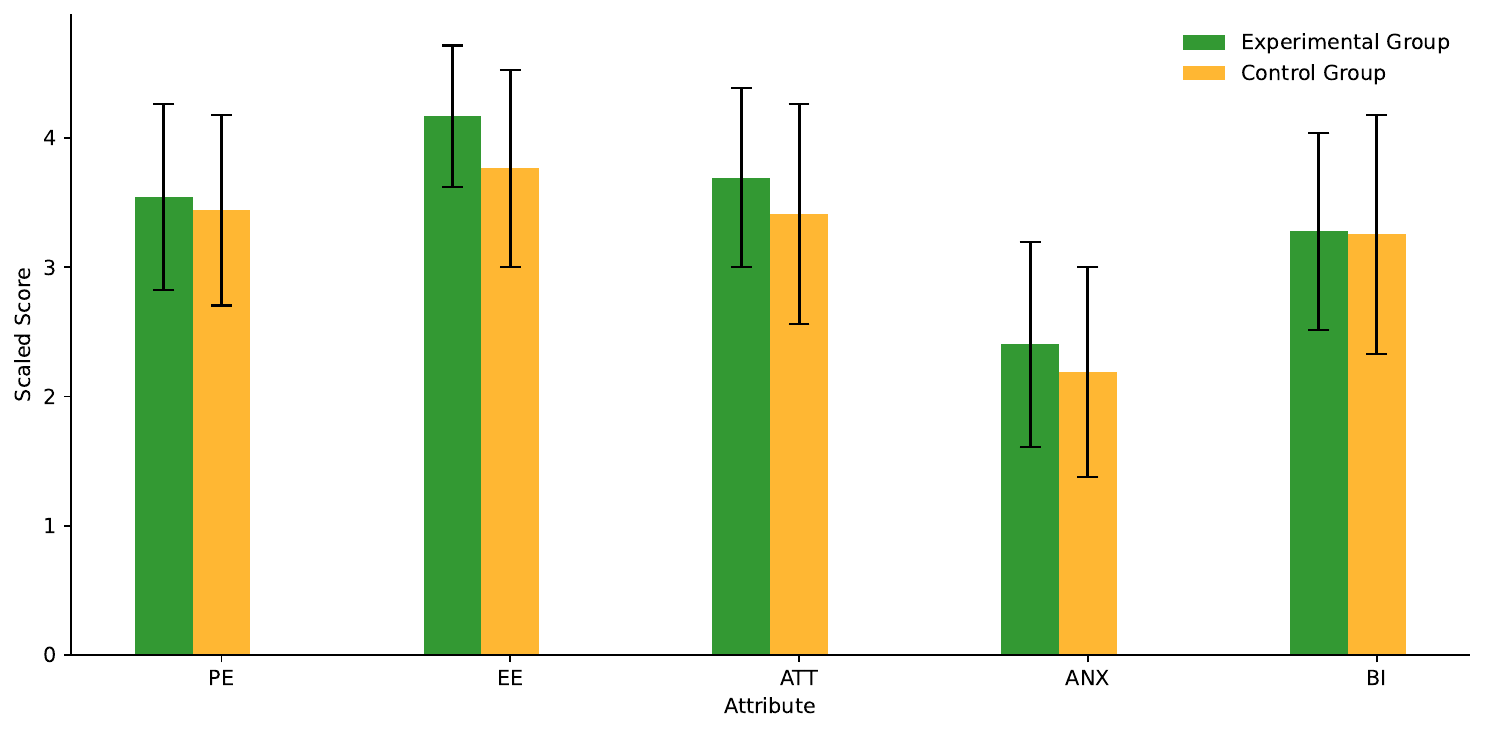}
    \caption{Scaled average values for measuring technology acceptance and use (UTAUT) across attributes: PE (Performance Expectancy), EE (Effort Expectancy), ATT (Attitude Toward using Technology), ANX (Anxiety), and BI (Behavioral Intention to use the system).}%
    \label{fig:UTAUT}%
    \Description{The bar graph compares the scaled average values for technology acceptance and use (UTAUT) attributes between the experimental and control groups, covering five attributes: Performance Expectancy (PE), Effort Expectancy (EE), Attitude Toward Technology (ATT), Anxiety (ANX), and Behavioral Intention (BI). For PE, the experimental group scored 3.54 (SD: 0.72), while the control group scored 3.44 (SD: 0.74). In EE, the experimental group had a higher score of 4.17 (SD: 0.55) compared to the control group’s 3.76 (SD: 0.76). For ATT, the experimental group scored 3.69 (SD: 0.69), slightly higher than the control group's 3.41 (SD: 0.85). In terms of ANX, the experimental group scored 2.40 (SD: 0.80), compared to the control group's 2.19 (SD: 0.81). Lastly, for BI, the experimental group scored 3.28 (SD: 0.76) compared to the control group's 3.25 (SD: 0.92).}
  \end{subfigure}

  \caption{Histograms showcase UX and technology acceptance and use, measured using UEQ and UTAUT questionnaires, respectively. Standard deviations are represented as error bars. No significant main effect was found between the experimental and control groups regarding the introduction of a new type of technology (i.e., \proto).}
  \label{fig:UEQ and UTAUT}

\end{figure}
To determine if the introduction of a new technology affected the ideation process from VBD, we analyzed self-reported data on participants' UX and technology acceptance from both the experimental and control groups, as shown in Fig. \ref{fig:UEQ and UTAUT}. We conducted one-way ANOVA and Kruskal-Wallis H tests for each attribute pair. The null hypothesis (H0) for these statistical tests assumed that there was no significant main effect between the two groups regarding attributes from UX and technology acceptance and use, meaning that the self-reported perceptions in both groups were the same. For the UEQ which measures UX (see Fig. \ref{fig:UEQ}), the analysis revealed no significant difference in the attractiveness attribute between the experimental group that used \proto and the control group as a baseline (ANOVA \(F_{1, 33} = .386\), \(p = .538\)). Similarly, comparisons of the other UEQ attributes—perspicuity (ANOVA \(F_{1, 33} = 1.208\), \(p = .332\)), efficiency (ANOVA \(F_{1, 33} = .008\), \(p = .944\)), dependability (ANOVA \(F_{1, 33} = 0.200\), \(p = .665\)), stimulation (ANOVA \(F_{1, 33} = 0.376\), \(p = .553\)), and novelty (ANOVA \(F_{1, 33} = 1.639\), \(p = .345\))—between the experimental and control groups also retained the null hypothesis (H0). Thus, \textbf{all six UEQ attributes collected from the experimental group using \proto measured UX has the same results as in the control group}. 

Additionally, as shown in Fig. \ref{fig:UTAUT}, the non-parametric Kruskal-Wallis test revealed that the PE attribute (performance expectancy) from UTAUT failed to reject the null hypothesis (\(x^2(1) = .003\),  \(p = .960\)), indicating no significant difference in performance expectancy between the groups. The mean rank scores were 17.92 for the experimental group, 18.09 for the control group. Similarly, the attributes of EE (effort expectancy) (ANOVA \(F_{1, 33} = 1.413\), \(p = .081\)), ATT (attitude toward using technology) (ANOVA \(F_{1, 33} = .699\), \(p = .287\)), ANX (anxiety) (ANOVA \(F_{1, 33} = .391\), \(p = .442\)), and BI (behavioral intention) (ANOVA \(F_{1, 33} = .004\), \(p = .938\)) also retained the null hypothesis. As such, \textbf{all attributes for measuring technology acceptance and use retained the null hypothesis between the two groups. These findings indicate that our experimental \proto did not introduce any negative effects on UX or technology acceptance and use compared to the normal ideation process in VBD (control)}.

\section{Discussion}
In this study, we conducted an A/B test to evaluate the impact of our \proto on the ideation process for VBD. Participants were assigned two sub-tasks and asked to generate as many design ideas as possible related to the provided contexts. Our findings indicate that \proto significantly enhanced participants' performance in terms of the flexibility and originality of their final ideation outputs compared to the baseline. Additionally, participants using \proto demonstrated greater engagement in decision-making, as evidenced by eye-tracking data, and there was a strong positive correlation between the number of ideas and words generated with \proto and the overall quality of their ideation. Furthermore, our findings suggest that the introduction of \proto did not negatively impact user experience or technology acceptance.
\subsection{Increased Flexibility and Originality in Divergent Thinking}
To measure creativity during ideation, Divergent Thinking is a well-established method supported by both theory and practice in prior studies \cite{ideation_runco_2010, runco_problem_1988, runco_divergent_2012}. In this study, we adopted this proven approach to investigate how a new tool (\proto) incorporating emerging technologies can enhance ideation within a design context involving videos. Our first research question (RQ1) explores the impact of \proto on ideation outcomes. To address this, we collected Divergent Thinking data from our study and had three independent graders with an "internal consistency" check to evaluate the quality of ideation, following principles outlined in well-established literature \cite{guilford_creativity_1950}. At the outset, we reviewed how ideation is understood and measured in the literature. For instance, fluency is used to assess the productivity of ideation, while flexibility indicates diverse ideas across different conceptual categories. Originality is defined by the novelty or rarity of ideas within a given task \cite{guilford_creativity_1950}. Our results show that participants in the experimental group, supported by \proto, received higher ratings in flexibility and originality compared to the control group. \textbf{This suggests that with \proto' assistance, the ideation process generated more multifaceted and novel ideas} \cite{runco_divergent_2012}. Specifically, the trait of flexibility could improve professional practitioners' understanding of tasks  (e.g., the usability of an artifact) and decision-making in design projects (e.g., plans for improvement) \cite{akin_creativity_1994}. Whereas originality, on the other hand, not only strongly correlates with innovation but also reflects the quality of authenticity and integrity of creative tasks \cite{guetzkow_what_2004}. Similarly, other studies concluded that ideation from industrial design tasks should consider three key aspects: "functional value", "aesthetic value" (e.g., visual form), and "originality value" \cite{christensen_dimensions_2016}. Our study showed that the prototype notably improved outcomes in two of these aspects—flexibility and originality. As such, the use of \proto enhanced the variety and novelty of ideas in creative VBD tasks.

\subsection{Greater Engagement in Ideation and Positive Correlation Between Interaction History and Performance}
We recognize that the final outcome of ideation (i.e., Divergent Thinking) was partially enhanced by the prototype. To explore further, we sought to understand how our prototyped \proto influenced the ideation processes in design tasks (RQ2). We began by measuring participants' eye movements during the tasks and observed an increased in pupil dilation in the experimental group compared to the control group. Previous studies have shown that dynamic changes in pupil dilation are associated with high-level cognitive processing \cite{hoeks_pupillary_1993}. Since the study was conducted in a stable lighting environment, the observed increase in pupil dilation indicates that participants voluntarily engaged in deeper, high-level decision-making prompted by the recommendations generated by \proto \cite{kang_pupil_2014}. Furthermore, the observed increase in gaze fixation duration and faster saccade speed in the experimental group suggests that participants were more engaged in the tasks compared to the control group \cite{colombo_individual_2014, hoffman_role_1995, spering_eye_2022}. Supported by existing literature, longer gaze fixation duration and quicker saccadic movements typically indicate higher levels of focus and cognitive engagement \cite{irwin_fixation_2004, zhao_eye_2012}. This may also suggest that our \proto captured participants' attention more effectively within the design task context compared to the traditional practice without additional helps in the control group. Similarly, the observed lower blink rate in the experimental group suggests that participants showed greater emotional interest in the generated content, which in turn increased their focus and engagement with the provided design use case \cite{maffei_spontaneous_2019}. A high level of work engagement has also been shown to lead to more positive and improved work performance \cite{kim_relationship_2013, coffrin_visualizing_2014}. In this way, \textbf{participants from the experimental group took the design-specialized advice and engaged in more iterative reflection in the ideation processing}. 

Following the eye-tracking measurements, we 
conducted a follow-up analysis of the chat logs. We examined how the \proto' responses influenced the interactions and how these exchanges correlated with the quality of ideas produced during human divergent thinking.

As noted in Section \ref{sec:chatlog}, many participants engaged with \proto to seek inspiration and guidance for potential design improvements based on the video context. The video comprehension function from \proto augmented the case debriefing process to allow participants to bypass the need to introduce the design case from scratch. Instead, participants could directly propose questions about both general and specific contents from the videos. The generative responses then effectively addressed the topics at hand and enabled participants to continue the conversation. Interestingly, participants also treated \proto as a companion in their design tasks. They often utilized its contextual understanding from a design professional's perspective to seek confirmation about the use cases or video contents. Upon receiving positive feedback, participants became more intrigued and confident which lead to deeper insights during the Divergent Thinking phase. Additionally, some participants incorporated their personal perspectives into the questions and findings they sought to confirm. This likely reflects the nature of design work, which is often driven by emotional and personal sentiments.

In addition, subsequent correlation tests reveal several strong and positive relationships between the words and ideas generated in chat logs and the quality of ideation in Divergent Thinking tasks. This suggests that participants' ideation holds positive relationship with the assistance from \proto. \textbf{Consequently, the ideation phase is likely to be enhanced by richer contents from generative answers in \proto}. Similarly, prior research has demonstrated that well-structured instructions in design tasks can play a significant role in eliciting higher levels of originality and fostering a broader range of ideation among practitioners \cite{runco_instructional_1991}. Additionally, we observed a positive correlation between the length of the questions asked by participants and the originality of their ideation. This suggests that the quality and quantity of the generated answers may be influenced by the level of detail in the query input. This finding brings our attention on the necessity of ensuring that human practitioners provide more detailed requests, clearly explaining their needs in the context of the current circumstances in future studies.

\subsection{No Decline in UX or Technology Acceptance and Use with the Introduction of New Technology} When introducing new technology into existing practices, practitioners may struggle with the adaptation process. Technostress, for example, is a phenomenon where individuals are unable to work with new information and communication technologies (ICT) in their work \cite{tarafdar_impact_2007}. This difficulty can lead to a decrease in productivity and creativity \cite{chandra_does_2019}. Previous literature has shown that discomfort with newly introduced tools often manifests as a decline in UX and in the ratings of technology acceptance and usage \cite{khlaif_impact_2023, tarafdar_impact_2010, tarafdar_technostress_2015}. Such a decline can potentially lead to ineffective use of the new technology and mismeasurement of its actual functionality. Given that the \proto integrates emerging ICT components, we are particularly interested in understanding whether the prototype affects UX and technology acceptance and use scores compared to the baseline (RQ3). In Section \ref{sec:uxandtam}, the analysis of self-reported scores from two separate questionnaires revealed no significant differences between the experimental and control groups. This suggests that participants in both groups exhibited similar levels of task satisfaction and willingness to accept and use the prototype. \textbf{As such, the proposed \proto did not negatively impact the normal design ideation experience and did not alter the original use and acceptance of the technology}. Additionally, while we observed lower ratings for certain attributes, such as perceived dependability, stimulation, and novelty within the user experience, these variations do not impact our overall findings of no significant difference in attribute scores. This may be attributable to individual attitudes towards the selected design scenarios, as design is inherently influenced by sentiments and emotions. We anticipate that future studies involving different VBD use cases may yield higher scores, though the pattern of results is expected to remain consistent.

\subsection{Limitations and Future Work}
While \proto shows significant potential for enhancing ideation in VBD, several limitations warrant further investigation. In informal post-experiment discussions, some participants expressed concerns around transparency and trust when using LLMs in creative processes. One of the primary challenges identified is the risk of "hallucinations," a common issue in AI-driven tools where models provide convincing yet incorrect information \cite{ji_survey_2023, lin_truthfulqa_2022}. This may increase confidence in creative tasks but can also lead to biased or flawed outcomes \cite{parasuraman_complacency_2010}. To mitigate this risk, we integrated the RAG mechanism \cite{lewis2020retr} into \proto. According to prior literature, RAG helps address the issue of generating inaccurate information by enabling the system to retrieve and incorporate task-centric, contextual-relevant, and factual-grounded content \cite{shuster_retrieval_2021}. In the future work, we aim to further enhance \proto’ transparency by integrating more interpretable outputs, such as providing citation links to credible literature in answers \cite{li_citation-enhanced_2024} which allow designers to trace the rationale behind generated suggestions.

Another limitation is the need to test \proto across a broader range of VBD use cases. While \proto proved effective in assisting design ideation within the two specific contexts of cooking and construction, real-world applications involve a much wider diversity of design tasks that may demand more flexible tools and an expanded knowledge base. In this study, we predefined the design books for \proto’s knowledge repository based on selections made by an independent committee to align with the study’s tasks. However, future work could allow designers to personalize the knowledge base by selecting and uploading their own domain-specific resources through a non-programmer-friendly interface. For example, platforms like AnythingLLM\footnote{\url{https://anythingllm.com/} (last accessed: \today)}  enable users to choose their own LLM models and indexed documents which could potentially offer a more tailored and flexible approach to ideation assistance. Furthermore, our current implementation of ideation assistance offers a fixed level of support to all users. However, design ideation is a highly individualized process, with varying needs for inspiration and suggestions based on the designer's experience \cite{jacob_design_2007, jacob_edit_2007}. To address this variability, we allowed participants in the study to critically consider their dependability of the assistance according to their own preferences, giving them the freedom to choose which aspects of ideation assistance to utilize and what to record in the Divergent Thinking process. The consistent level of support was maintained to ensure a fair comparison and to isolate \proto' impact on ideation. In future iterations, we could consider \proto as a product and implement a tunable feature that allows users to adjust the level of "helpfulness" in guiding the design ideation process. We expect this would enable designers to control the amount of information provided according to their needs and makes the tool more responsive to individual preferences.

\section{Conclusion}
The advancement of generative AI has substantially transformed human work in recent years. In VBD design, there remains an urgent need to reduce the burden of manual video analysis and accelerate professional ideation. Prior research across multiple disciplines has demonstrated efforts to harness the power of generative AI to augment design ideation. In this paper, we present \proto, a prototype that elevates ideation assistance for VBD to a higher level. Utilizing advanced techniques from generative AI, our \proto can automatically extract information from videos, integrate with professional design guidelines from indexed literature, and provide design- and case-centric recommendations to inspire designers. Our findings demonstrate that \proto significantly improves ideation outcomes in terms of flexibility and originality in Divergent Thinking. Through cognitive monitoring via eye-tracking and chat log analysis, we observed increased engagement in design ideation when using \proto. Furthermore, assessments of UX and technology acceptance and use indicated that the introduction of this tool did not contribute to increased stress and ensures there will be a smooth integration into the existing VBD workflow in future.

\bibliographystyle{ACM-Reference-Format}
\bibliography{main.bib}

\appendix

\begin{quote}
    \framebox{\parbox{\dimexpr\linewidth-2\fboxsep-2\fboxrule}{
\textit{You will be shown two videos. Your task is to analyze the videos and pinpoint processes or methods that could be enhanced. Focus on the activities and consider alternative tools, interactions, or contextual improvements. Generate and write out as many ideas as possible. You are encouraged to think out loud. \\ \\ (Please use the provided chatbot to assist you. This tool offers insights and suggests improvements based on the video content. Type your questions or thoughts into the chatbot and use its responses to enhance your ideation. For example, ask, "How can the process shown in the video be improved?") \\ \\ You will have 15 minutes to engage with each video. Please use your time effectively and document as many ideas as possible. Please note that videos do not have sound. You will be notified after 12 minutes of the time. \\ \\ When you are ready to proceed press the "Start" button and the arrow "$\rightarrow$" on the bottom right side of the screen.}
}
}
\captionof{floatquote}{Instructional text displayed in the Note-taking Space in Fig. \ref{fig:UI}. Text within parentheses (the second paragraph) was shown only to participants in the experimental group with access to \proto.}
\label{instruction text}%

\end{quote}

\begin{quote}
    \framebox{\parbox{\dimexpr\linewidth-2\fboxsep-2\fboxrule}{
\begin{itemize}
\item \textit{Flexibility: Each comprehensive idea which portraying the purpose and functionality in sufficient detail to be understandable gives a +1 point. }
\item \textit{Flexibility: Give a +1 point for each new domain/subdomain is spotted based on the ideation context across all participants. }
\item \textit{Originality: A grade based on the statistical infrequency of ideas measured on a 7-point Likert scale.}
\end{itemize}
}
}
\captionof{floatquote}{Predetermined criteria based on Guilford’s study \cite{guilford_creativity_1950} for evaluating fluency, flexibility, and originality in divergent thinking texts by independent raters.}
\label{divergent thinking criteria}%

\end{quote}

\end{document}